\documentclass[twocolumn,preprintnumbers,superscriptaddress,pre]{revtex4-2}
\usepackage{graphicx, subfigure}
\usepackage{amssymb, amsmath,amssymb,amsfonts}
\usepackage{amsthm,mathrsfs,amsopn}
\usepackage{dcolumn}
\usepackage{bm}
\usepackage{color}
\usepackage[utf8]{inputenc}
\usepackage{tikz} % for plots, graphics
\usepackage{pgfplots}
\usepackage{comment}

\theoremstyle{plain} %% This is the default

\definecolor{fluorescentpink}{rgb}{1.0, 0.08, 0.58}

\begin{document}

\title{Complex contagion in social systems with distrust}

\author{Jean-François de Kemmeter}
\email{jean-francois.dekemmeter@unamur.be}
\affiliation{Department of Mathematics and naXys, Namur Institute for Complex Systems, University of Namur, Rue Graf\'e 2, B5000 Namur, Belgium}
\affiliation{Department of Mathematics, Florida State University, 1017 Academic Way, Tallahassee, FL 32306, United States of America}

\author{Luca Gallo}
%\email{gallol@ceu.edu}
\affiliation{Department of Network and Data Science, Central European University, Quellenstraße 51, 1100 Vienna, Austria}

\author{Fabrizio Boncoraglio}
%\email{fabrizio.boncoraglio@gmail.com}
\affiliation{Department of Applied Science and Technology, Politecnico of Turin, Corso Duca degli Abruzzi 24, 10129 Turin, Italy}
\affiliation{Scuola Superiore di Catania, Via Valdisavoia 9, 95123 Catania, Italy}

\author{Vito Latora}
%\email{v.latora@qmul.ac.uk}
\affiliation{School of Mathematical Sciences, Queen Mary University of London, London E1 4NS, UK}
\affiliation{Department of Physics and Astronomy,  University of Catania, 95125 Catania, Italy}
\affiliation{INFN Sezione di Catania, Via S. Sofia, 64, 95125 Catania, Italy}
\affiliation{Complexity Science Hub Vienna, A-1080 Vienna, Austria}

\author{Timoteo Carletti}
\email{timoteo.carletti@unamur.be}
\affiliation{Department of Mathematics and naXys, Namur Institute for Complex Systems, University of Namur, Rue Graf\'e 2, B5000 Namur, Belgium}

\begin{abstract}
Social systems are characterized by the presence of group interactions and by the existence of both trust and distrust relations. 
Although there is a wide literature on signed social networks, where positive signs associated to the links indicate trust, friendship, agreement, while negative signs represent distrust, antagonism, and disagreement, very little is known about the effect that signed interactions can have on the spreading of social behaviors when, not only pairwise, but also higher-order interactions are taken into account. 
In this paper we focus on processes of complex contagion, such as the adoption of social norms or the diffusion of novelties, where exposure to multiple sources is needed for the contagion to occur. 
Complex contagion has been recently modeled, at a microscopic scale, by higher-order networks, such as simplicial complexes, which allow transmission to happen not only through the links connecting pair of nodes, but also in group interactions, namely over simplices of dimension larger or equal than two. Here, we introduce a model of complex contagion on signed simplicial complexes, and we investigate the role played by trust and distrust on the dynamics of a social contagion process. 
The presence of higher-order signed structures in our model naturally induces new infection and recovery mechanisms, thus increasing the richness of the contagion dynamics. Through  numerical simulations and analytical results in the mean-field approximation, we show how distrust determines the way the system moves from a state where no individuals adopt the social behavior, to a state where a finite fraction of the population actively spreads it.
Interestingly, we observe that the fraction of spreading individuals displays a non-monotonic dependence on the average number of connections between individuals. 
We then investigate how social balance affects social contagion, finding that balanced triads have an ambivalent impact on the spreading process, either promoting or impeding contagion based on the relative abundance of fully trusted relations.
Our results shed light on the nontrivial effect of trust on the spreading of social behaviors in systems with group interactions, paving the way to further investigations of spreading phenomena in structured populations. 
\end{abstract}
\maketitle

\noindent 

\section{Introduction}
\label{sec:intro}
% what is a spreading process, what is a complex contagion
Contagion is the process through which a disease, an opinion, a behavior or a technological innovation  spreads in a population of interacting individuals \cite{barrat2008dynamical}. 
While some spreading processes only require a single exposure 
for a contagion to occur,  others are only triggered by  multiple exposures. 
In the first case, the spreading process is referred to as \textit{simple contagion}, while in the latter case it is known as \textit{complex contagion} \cite{centola2007complex}. 
The spread of an infectious disease in a population is a typical example of a simple contagion \cite{pastor2015epidemic}, while the propagation of a social behavior, such as the adoption of a social norm, of an opinion, or even a novel item
can be characterized either as a simple \cite{hill2010infectious} or a complex contagion \cite{lehmann2018complex}. 
In particular in social contagion there are cases where the individuals/agents of a social system can either accept or refuse transmission, based on the type of their social interactions, instead of just passively receive it as in the case of a virus \cite{kiss2017mathematics}. 
%

% complex contagion is shaped by the network of interactions
The spread of a social behavior in a population is utterly determined by the precise patterns of interactions among individuals. 
Usually, the structure of contacts in a social system is mathematically modeled as a graph, whose nodes represent the individuals, the social agents, while the edges encode for their pairwise interactions  \cite{latora2017complex}.
Without surprise thus, the contagion process is shaped by the structure of such graph, 
for instance by its modular organization, the existence of short paths, weak ties or influential nodes~\cite{ghasemiesfeh2013complex,malliaros2016locating,pei2013spreading,ruan2015kinetics}.
Complex contagion on networks has been notably analyzed using threshold models, where a susceptible individual becomes infected if a sufficiently large fraction of its neighbors is infected~\cite{lee2017social,wang2016dynamics}. 

% higher-order networks
Although network science provides powerful tools to model dynamical processes in social systems, the network approach has strong limitations in dealing with group interactions. 
The latter are indeed a fundamental ingredient of social systems and result to be a main driver for social contagion~\cite{iacopini2019simplicial}.
As a consequence, last years have seen a growing interest in higher-order networks, namely  mathematical structures generalizing complex networks and accounting for group interactions~\cite{battiston2020networks,battiston2021physics,majhi2022dynamics,bianconi2021higher,bick2021higher}. 
Recent works on social contagion on higher-order networks have demonstrated that peer pressure and group reinforcement can induce novel phenomena, such as the appearance of abrupt phase transitions and bistability~\cite{iacopini2019simplicial,de2020social,landry2020effect,chowdhary2021simplicial,barrat2022social,st2022influential,ferraz2023multistability}.
One aspect that still has not been considered and explored exhaustively is the role that trust/distrust can have on the diffusion of a novel idea or item, or in the adoption of a social norm. 
Trust among individuals is a key driver of contagion in social systems: agents are more prone to adopt a novel behavior if it comes from trusted sources. 
Moreover, infected agents, i.e., agents having adopted a given item, when they interact with other infected agents that they do not trust, can decide to stop adopting the item. 
The rationale behind the latter process is that two agents that distrust each other would not be inclined to share the same item. 
%Some works have already started to study the effect of trust and distrust on simple and complex contagion over a social network. 
Some researches have already started to investigate the impact of trust on social contagion and epidemic spreading on  networks~\cite{saeedian2017epidemic,unicomb2018threshold,li2021dynamics}, but no work has yet considered contagion processes on signed higher-order networks.

\medskip
In this paper we propose a model of complex contagion on signed higher-order networks, which allows to investigate the impact of trust and distrust among individuals on the spread of contagion in a social system, by properly taking into account interactions in groups of two or more individuals at the microscopic level. 
Our model is inspired by the model of complex contagion on simplicial complexes proposed in Ref.~\cite{iacopini2019simplicial}.   
Indeed, as the underlying structure of the social system we consider a simplicial complex instead of a network. 
A simplicial complex is a made not only by nodes and by links 
connecting pairs of nodes, and 
representing pairwise interactions between  individuals, but also by more complex objects, which  represent interactions in groups of more than two individuals. 
We then extend the model of Ref.~\cite{iacopini2019simplicial} by introducing positive and negative signs over the links of the simplicial complex. 
It is however a common experience that the trust/distrust relations existing between two individuals can change once the two individuals are interacting into a group of larger size. 
For this reason we propose and analyze two variants of our social contagion model. 
In the first version of the model we assume that the signs of the pairwise interactions remain the same also in higher-order interactions. 
In the second one, we relax such an assumption allowing  agents to exhibit pairwise trust/distrust relations different from those experienced in groups involving more than two individuals. Namely, in order to determine the distribution of signed relations in a group interactions, we recur to {\em social balance} theory~\cite{heider1946attitudes,harary1953notion,leskovec2010signed,zheng2015social,kirkley2019balance}. 
According to this theory a group of three individuals, a triad, is balanced when there is an odd number of relations among them. Thus we have two distinct balanced  configurations that can be interpreted either as ``the friend of my friend is my friend'' in case of three trust relations, or ``the enemy of my enemy is my friend'' in presence of a single trust relation. 
Both of these configurations ensure the consistency of the members' attitude with respect to each other. 
At contrast, a triad in which there is an even number of trust relations leads unavoidably to some kind of frustration among the individuals in the triad as it implies that the ``friend of my friend is my enemy'' or that ``the enemy of my enemy is my enemy'', both cases being unwanted by its members. For this reason we expect these types of triads to appear less frequently. 

% the results 
With the first version of our model, we show how the level of distrust in the social network determines the nature of the transition, i.e., continuous or discontinuous, from a state where no individual adopts the social behavior, i.e., disease-free state, to a regime where a fraction of agents spread it, i.e., to an endemic state.
Particularly, we show that the size of the endemic state can be a non-monotonic function of the average node degree, depending of the level of distrust in the population.
With the second version of the model, we analyze the ambivalent role of social balance, showing that simplicial complexes where balanced triads are overrepresented compared to the random case can either promote or impede contagion.
We corroborate our findings by comparing the mean-field predictions to stochastic simulations performed by using a Gillespie algorithm~\cite{gillespie}.
In the framework of social balance theory, namely the second proposed model, we show that a structurally balanced network - a network in which balanced triads are overrepresented compared to the random case - can either promote or impede contagion.  
In particular, we find that the abundance of triangles with three positive signs enhances the emergence of a bistable regime and thus lets the system easily switch to an endemic state. We also show that the other balanced configuration, i.e., two negative signs, induces instead the opposite behavior. Similar results are also presented when investigating the role of unbalanced triangles.

% the structure of the paper
The paper is organized as follows. In Section II, we introduce our model of complex contagion on signed simplicial complexes. In Section~\ref{sec:meanfield}, we 
derive the mean-field equations of the model, while  
in Section~\ref{sec:statdens}, we determine the equilibrium solutions of the system and examine their stability, assuming a random distribution of edge signs, namely under the assumption of the first model. In section~\ref{sec:sbt}, we introduce the second version of the model, and we investigate how structural balance affects contagion. We then conclude with some future perspectives. 

%NB: More technical features of the model were erased from the introductory part and are going to be integrated in the next section "The Model" 
%Better "Model of social contagion with trust and distrust"
\section{Modeling social contagion with trust and distrust}
\label{sec:themodel}
% introducing signed simplices
In the context of pairwise interactions, trust and distrust relations between individuals are represented as links associated with signs, respectively positive or negative.
Inspired by this, to model group interactions with trust or distrust we will use signed simplices. 
A $q$-simplex, $\sigma$, is a collection of $q+1$ distinct nodes. With such a definition, a $0$-simplex is a node, a $1$-simplex is a link, corresponding to a dyadic interaction, a $2$-simplex is a triangle, representing an interaction among three individuals, and so forth. 
To model trust and distrust, each node in a $q$-simplex, $q\geq 1$, is endowed with a positive or negative sign toward each of the remaining $q$ nodes of the simplex.
For simplicity, we assume (dis)trust relations to be reciprocal.
Let us observe that in real social systems the existence of a (dis)trust relation between two individuals in a pairwise interaction does not imply that the same (dis)trust relation exists when a third person is involved in the interaction.

% introducing the rules of the model
In our model, individuals are assumed to be either in an infected state, i.e., they are spreading a given social behavior, or in a susceptible state, i.e., they are not.
Note that we will hereafter keep using the vocabulary of epidemic processes. 
In particular, we will denote endemic state the condition of the system where a positive fraction of individuals are infected, i.e., they have adopted and spread the social behavior.
Also, we will refer to the threshold above which an endemic state emerges as epidemic threshold. 
However, compared to epidemic spreading, in our model we assume that individuals can adopt but also refuse to adopt the social behavior under study, namely because they can change their mind by interacting with others. 
This is impossible in the case of disease spreading.

\begin{figure}[t]
	\includegraphics[scale=.3]{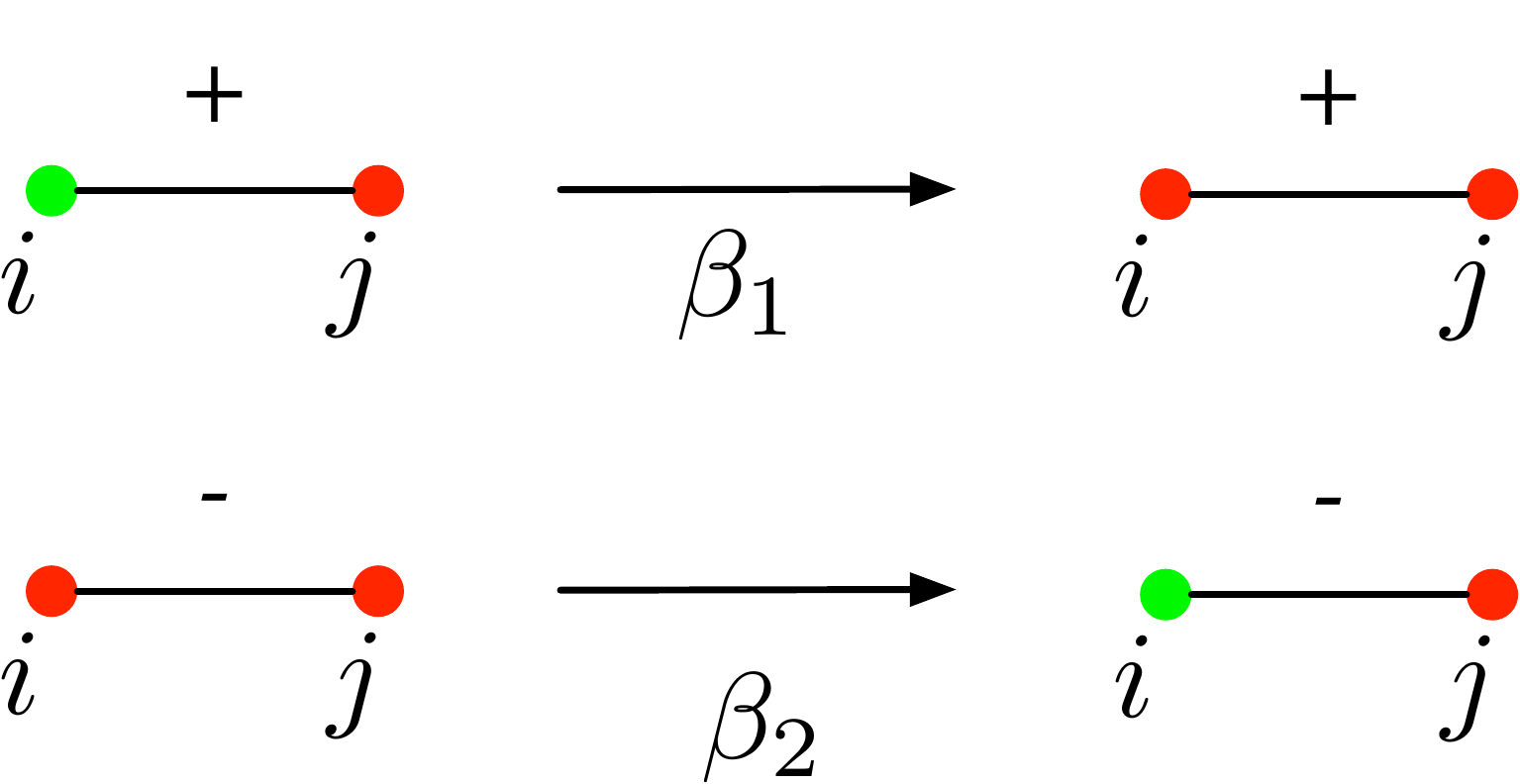}
	\caption{\textbf{Infection and recovery in signed pairwise interactions}. A susceptible agent (green) gets infected (red) by a trusted infected agent (top panel) with a rate $\beta_1>0$. An infected agent recovers,  when in contact with a distrusted infected agent (bottom panel), with a rate $\beta_2>0$. In all the remaining pairwise configurations the agent will not change its state.}
	\label{fig:case0}
\end{figure}
% 1-body and 2-body processes
In our model, we assume that infection and recovery processes can take place at three distinct levels, namely at the level of single individuals, i.e., $0$-simplices, at the level of pairwise contacts, i.e., $1$-simplices, and also at the level of three-body interactions, $2$-simplices.
To present our results in a simple and direct way, we limit our analysis here to simplices up to order $q=2$.
This assumption is motivated by the fact that larger interacting groups are less abundant compared to smaller groups~\cite{cencetti2021temporal}.
However, our modeling approach can be straightforwardly extended to larger group sizes~\cite{de2020social}.
At the level of $0$-simplices, we only have recovery processes, with infected individuals coming back to the susceptible state with a rate $\mu$, namely they stop spreading after an average time $1/\mu$.
At the level of $1$-simplices, two processes are at play, an infection process and a recovery one.
For the infection process, a susceptible individual gets infected with a rate $\beta_1>0$ when interacting with a trusted infected individual (see top panel of Fig.~\ref{fig:case0}). 
For the recovery process, if an infected individual interacting with a distrusted infected individual becomes susceptible with a rate $\beta_2>0$ (see bottom panel of Fig.~\ref{fig:case0}).
Two-body recovery processes have not been extensively adopted, with a few exceptions~\cite{iacopini2020multilayer}, and in our case the idea behind it is that individuals that do not trust each other are less willing to adopt and spread the same social behavior.

\begin{figure}[t]
	\includegraphics[scale=.23]{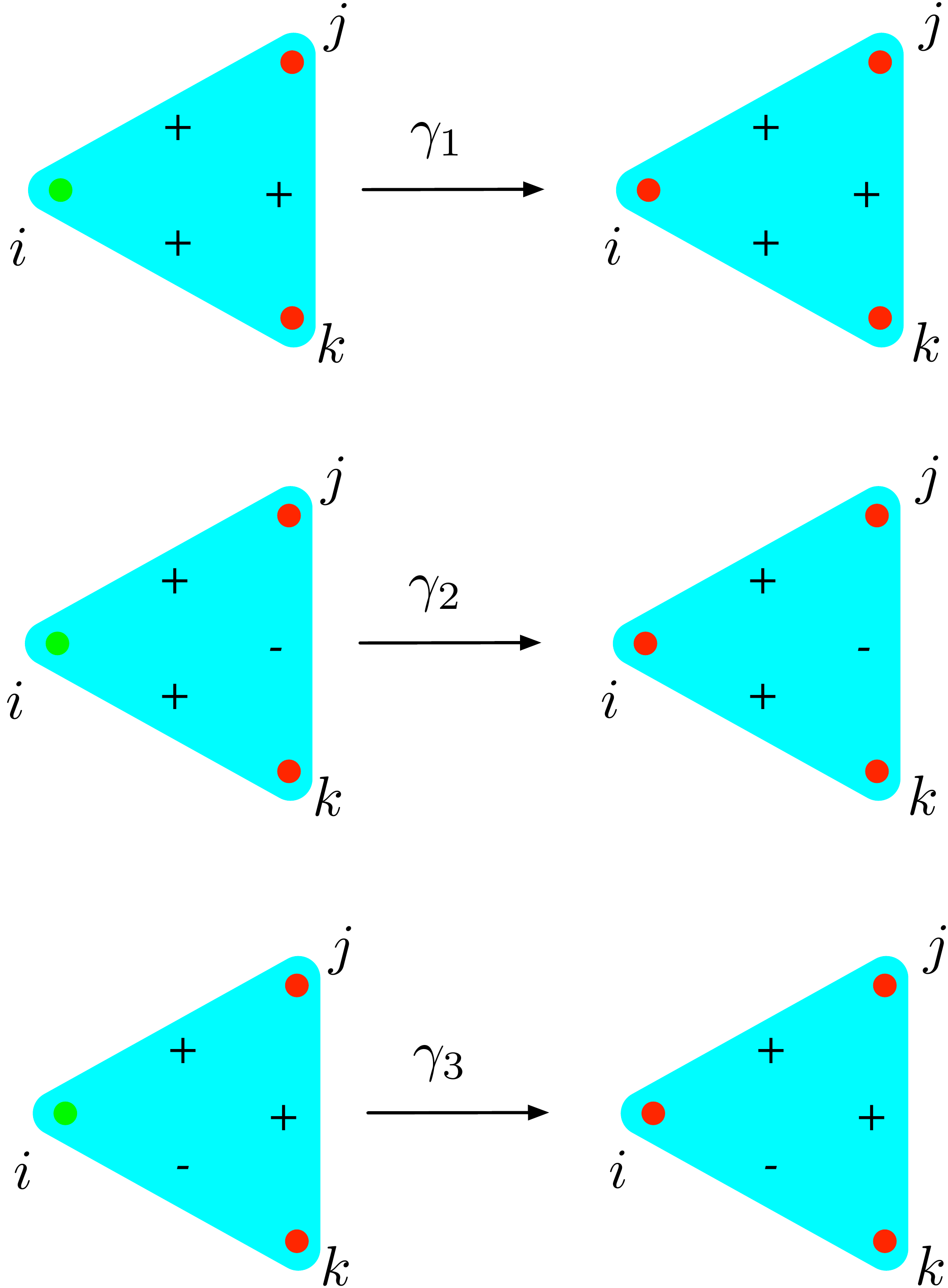}
	\caption{\textbf{Infection in groups of size $3$ with signed relations}. A susceptible agent (green) becomes infected (red) because of an interaction in a group of size $3$ (cyan triangle). In the top panel the agents experience three trust relations, in the middle panel there is distrust between the two infected nodes $j$ and $k$,  and in the bottom panel there is distrust between $i$ and $k$, i.e., a susceptible and an infected node. In all three cases the susceptible agent will get infected but with rates $\gamma_1\geq \gamma_2\geq \gamma_3\geq 0$.}
	\label{fig:case1}
\end{figure}

% 3-body processes
In the case of three-body interactions, we consider five different processes, namely three types of infections and two types of recoveries.
An individual can change state by interacting with two infected individuals, with a rate that depends on the trust/distrust relations existing among the three agents. 
This comes from the assumption that, when interacting in group, each agent has full knowledge of all the trust/distrust relations existing among the group members.  
Note that this is a reasonable assumption in the case of small groups.
Let us first consider the cases in which a susceptible individual, $i$, participates in a three-body interaction with two infected ones, $j$ and $k$ (see Fig.~\ref{fig:case1}). 

{\bf Case 1:}
If all the agents trust each other, i.e., there are three ``$+$'' signs in the $2$-simplex, then the susceptible individual $i$ becomes infected with a rate $\gamma_1>0$. 

{\bf Case 2:}
If $i$ trusts $j$ and $k$ but there is a distrust relation between $j$ and $k$, $i$ can still be infected, but with a positive rate $\gamma_2 \leq \gamma_1$. 
This comes from the idea that the distrust relation existing among $j$ and $k$ can negatively influence the propensity of $i$ to adopt the social behavior.

{\bf Case 3:}
If a susceptible agent $i$ trusts agent $j$ but distrusts agent $k$, while there is a trust relation between $j$ and $k$, then $i$ becomes infected with a positive rate $\gamma_3\leq\gamma_2$. 
As in the previous case, the presence of a distrust relation determines a decrease in the adoption rate of agent $i$, even more than in the case of the second mechanism, as agent $i$ directly experience distrust.

\begin{figure}[t]
	\includegraphics[scale=.23]{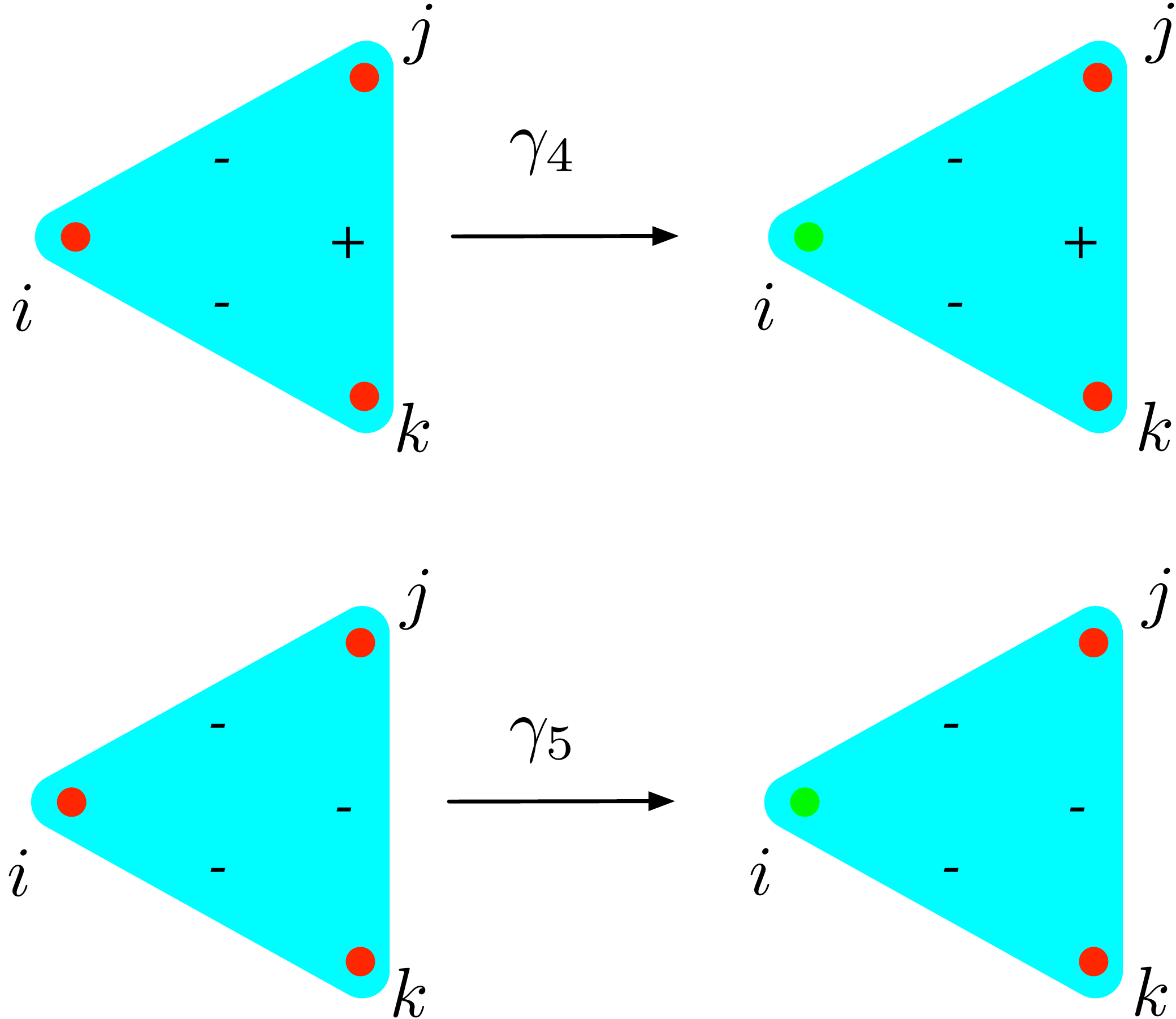}
	\caption{\textbf{Recovery in groups of size $3$ with signed relations}. An infected agent (red) becomes susceptible (green) because of the interaction in a group of size $3$ (cyan triangle). In the top panel agent $i$  experiences a distrust relation with $j$ and $k$, and at the same time perceives a trust relation between $j$ and $k$. In the bottom panel all the agents distrust each other. In both cases $i$ will become susceptible, in the former case with a rate $\gamma_4\geq0$, in the latter with a rate $\gamma_5\geq0$, such that $\gamma_4\geq\gamma_5$.}
	\label{fig:case2}
\end{figure}

We now consider the cases where three infected individuals, experiencing distrust relations, interact (see Fig.~\ref{fig:case2}).

{\bf Case 4:}
If an agent $i$ distrusts both $j$ and $k$, perceiving a trust relation between them, it will become susceptible with a rate $\gamma_4\geq0$.

{\bf Case 5:}
If the three agents distrust each other, one of them can still become susceptible, but at a smaller rate, i.e., $0\leq\gamma_5\leq\gamma_4$. 
In both cases, the rationale is that distrust can induce doubt, pushing the agents to stop adopting the social behavior.

So far, we have described the microscopic mechanisms, occurring at the level of $q$-simplices, underlying our model of social contagion with trust and distrust.
However, the macroscopic spreading of social behaviors within a population is determined by the precise patterns of interaction among individuals.
To model the complex structure of interactions in a real social systems, we rely on signed simplicial complexes. 
Given a set $\mathcal{V}=\{1,\dots,N\}$ of nodes, a simplicial complex $\mathcal{K}$ is a collection of signed simplices, with an inclusion requirement, meaning that for any simplex $\sigma\in\mathcal{K}$, all simplices $\tau\subset\sigma$ are also contained in $\mathcal{K}$.
Let us observe that the inclusion property means, for example, that if three people interact altogether, then each of them will also interact pairwise with the two others.
In the following, we will denote by $\eta\in[0,1]$ the fraction of $1$-simplices (the links) with a minus sign in the simplicial complex, i.e., the fraction of dyadic relations based on distrust. 
Let us observe that besides the epidemic parameters and the number of trust/distrust relations, our model of social contagion will also depend on the topological features of the simplicial complex representing the interactions among the individuals, e.g., the arrangement of trust/distrust relations, the average number of $1$ and $2$-simplices incident to a node, and so on. 
In the following we will focus on the impact of the distrust using mean-field equations and stochastic simulations. 

\section{Mean-field equations}
\label{sec:meanfield}
Let us now cast the microscopic mechanisms discussed so far into an ODE model for the fraction $\rho(t)$ of infected agents at time $t$. Working in a mean-field approximation that neglects $2$-nodes correlations, we get:
\begin{eqnarray}
\label{eq:main}
&\dfrac{d\rho}{dt}&=-\mu \rho+\beta_1 \rho(1-\rho)\ell_1-\beta_2\rho^2 \ell_0\notag\\
&+&\gamma_1\rho^2 (1-\rho)\tau_{3}+\frac{\gamma_2}{3}\rho^2 (1-\rho)\tau_{2}+\frac{2\gamma_3}{3}\rho^2 (1-\rho)\tau_{2}\notag\\
&-&\frac{\gamma_4}{3}\rho^3\tau_{1}-\gamma_5\rho^3\tau_{0}\, ,
\label{eq:drhodt}
\end{eqnarray}
where $\ell_1$ (resp. $\ell_0$) is the mean number of 
trust links, i.e. links with a ``$+$'' sign (resp. distrust links, i.e. with a ``$-$'' sign), which are incident in a node. We can write $\ell_1=(1-\eta) \langle k\rangle$, and thus $\ell_0=\eta\langle k\rangle$, where $\langle k\rangle$ is the average node degree, and $\eta$ the fraction of negative links existing in the social network. For $m\in\{0,1,2,3\}$, the mean number of triangles with $m$ trust relations, i.e., $m$ ``$+$'' signs, is denoted by $\tau_{m}$.

The different terms in the first line of Eq.~\eqref{eq:drhodt}  respectively model (from the left to right) the standard (single-node) recovery, the infection due to a pairwise interaction between a susceptible and an infected person trusting each other, and finally the recovery of an infected agent because of a dyadic interaction involving two distrusting infected agents (i.e., the processes described in Fig.~\ref{fig:case0}). The second line, again from left to right,  describes a contagion due to a triadic interaction involving two infected and one susceptible agent, all trusting each other (see top panel of Fig.~\ref{fig:case1}). A contagion occurring in a triadic group with two infected and one susceptible people, the latter trusting the two others who do not trust each other; the factor $1/3$ represents the fact that, focusing on one node, only one third of all triangles with two ``$+$'' and one ``$-$'' has the right configuration of signs (see middle panel of Fig.~\ref{fig:case1}). Finally, a contagion due to again a triadic interaction with two infected individuals and a susceptible one, where the susceptible person trusts one person but not the other one. The factor $2/3$ again is due to the fact that only a fraction $2/3$ of such triangles has the right signs setting (see bottom panel of Fig.~\ref{fig:case1}). The third line of Eq.~\eqref{eq:main} denotes the recovery of an infected person because of a triadic interaction involving three infected people, the former one distrusting the latter two, from which the factor $1/3$ follows. Then a recovery process involving three mutually distrusting infected people (see Fig.~\ref{fig:case2}).

Gathering together powers of $\rho$ we can rewrite Eq. ~\eqref{eq:main} as follows:
\begin{equation}
\label{eq:edo}
\frac{d\rho}{dt}=\rho \left(a +b \rho+c \rho^2\right)\, ,
\end{equation}
where
\begin{eqnarray*}
	a &=& -\mu +\beta_1 (1-\eta) \langle k\rangle\\
	b &=& - \beta_1 (1-\eta) \langle k\rangle -\beta_2 \eta\langle k\rangle+\gamma_1\tau_{3}+\frac{\gamma_2\tau_{2}}{3}+\frac{2\gamma_3\tau_{2}}{3}\\
	c &=& -\gamma_1\tau_{3} -\frac{\gamma_2\tau_{2}}{3}-\frac{2\gamma_3\tau_{2}}{3}-\frac{\gamma_4}{3}\tau_{1}-\gamma_5\tau_{0}\, .
\end{eqnarray*}

\section{Stationary densities: existence and stability}
\label{sec:statdens}

We are now interested in finding the stationary solutions of Eq.~\eqref{eq:main} and in studying their stability. This will allow us to investigate the asymptotic behavior of our system and to draw conclusion on the conditions leading to the emergence on an endemic state.
Let us start by assuming that the three agents taking part in a three-body interaction maintain the same trust and distrust relations they have in pairwise interactions, meaning  that the signs of the three-body interaction are directly ``inherited'' from the three underlying pairs. 
This will allow us to draw some preliminary conclusions on the role of trust on contagion. 
we will then relax such an hypothesis and present a more general setting in the following section. Under this working assumption, we can write
\begin{eqnarray*}
	\tau_{0}&=&\eta^3\langle k_\Delta\rangle\, , \tau_{1}=3\eta^2(1-\eta)\langle k_\Delta\rangle\\
	\tau_{2}&=&3\eta(1-\eta)^2\langle k_\Delta\rangle\text{ and }\tau_{3}=(1-\eta)^3\langle k_\Delta\rangle\, ,
\end{eqnarray*}
where $\langle k_\Delta\rangle$ is the average number of $2$-simplices incident to a node. 
We can thus determine the stationary solutions of the mean-field equations and their stability. Beside the disease free equilibrium $\rho_0^*=0$, that is always a solution, there can be other two solutions given by:
\begin{equation}
\rho^*_{\pm} = \frac{
	\Lambda - \tilde{\lambda}_1- \lambda_2 
	\pm \sqrt{ {(\Lambda- \tilde{\lambda}_1- \lambda_2)}^2
		-4(\Lambda + \tilde{\Lambda})(1-\tilde{\lambda}_1)}
}{
	2(\Lambda + \tilde{\Lambda})\, ,
}
\label{rhopm}
\end{equation}
if $(\Lambda- \tilde{\lambda}_1- \lambda_2)^2-4(\Lambda + \tilde{\Lambda})(1-\tilde{\lambda}_1)>0$ and the parameters are such that $\rho_{\pm}^*\in (0,1)$, where $\Lambda := \Lambda_1+\Lambda_2 + \Lambda_3$ and  $\tilde{\Lambda}:=\Lambda_4 + \Lambda_5$ with:
\begin{eqnarray}
\Lambda_1 &=& \frac{\gamma_1\tau_{3}}{\mu} \, , \Lambda_2 = \frac{\gamma_2\tau_{2}}{3\mu} \, , \Lambda_3 = \frac{2\gamma_3\tau_{2}}{3\mu}\, , \label{eq:Lambda123}\\
\Lambda_4 &=& \frac{\gamma_4\tau_{1}}{3\mu} \text{ and }\Lambda_5 = \frac{\gamma_5\tau_{0}}{\mu}\, , \label{eq:Lambda45}\\
\tilde{\lambda}_1 &:=& {\lambda_1 (1-\eta)} = \frac{\beta_1 \langle k \rangle}{\mu}  (1-\eta), \lambda_2 = \frac{\beta_2 \eta \langle k \rangle}{\mu}\, . \label{eq:lambda12}
\end{eqnarray}
We notice that, in the special case where all agents trust each other, i.e. when $\eta=0$, {we have $\tilde{\Lambda} = 0, \lambda_2 = 0$, $\Lambda=\lambda_\Delta:=\frac{\gamma_1 k_\Delta}{\mu}$ and $\tilde{\lambda}_1=\lambda_1:=\frac{\beta_1 \langle k \rangle}{\mu}$ and} we recover the fixed points obtained in~\cite{iacopini2019simplicial}. {In this case, it has been shown that the epidemic threshold is given by $\lambda_1=1$ and that the transition at the epidemic threshold is continuous when $\lambda_\Delta \leq 1$ and discontinuous otherwise; in that case the system shows bistability for an interval of values of $\lambda_1$, i.e., it admits as stable equilibria both the disease free state and the endemic state. In the following we show how the analysis can be extended to the general case where $0 < \eta \leq 1$.}

By rescaling time by $\mu$ and linearizing~\eqref{eq:edo} close to the equilibrium $\rho_0^*=0$, we can realize that the latter is stable if and only if $\tilde{\lambda}_1=\lambda_1 (1-\eta)<1$. Hence, we find that the epidemic threshold for the disease free equilibrium is given by $\lambda_1=1/(1-\eta)$. Above this threshold, the infection will always spread and reach a non-zero fraction of the population, while {below this threshold} the infection will die out unless the fraction of initially infected nodes is sufficiently large, i.e., above a critical mass. In that case nonlinear terms might sustain the disease. Crucially, we observe that the epidemic threshold increases as $\eta$ becomes larger (see top panel of Fig.~\ref{fig:rhostar_1}). This means that distrust can limit the spreading of the infection, as a large value of $\lambda_1$ is needed to initiate the contagion process.   
Let us now analyze the existence and the stability of the aforementioned fixed points $\rho_{\pm}^*$, as a function of $\tilde{\lambda}_1$. 
\begin{itemize}
	\item By looking at the definition of $\rho^*_\pm$ one can realize that if $0<\tilde{\lambda}_1<\lambda_{\mathit{crit}}$ then $\rho^*_\pm$ assume complex values, where  $\lambda_{\mathit{crit}}=-(\lambda_2+\Lambda+2\tilde{\Lambda})+2\sqrt{(\Lambda+\tilde{\Lambda})(1+\lambda_2+\tilde{\Lambda})}<1$ as one can observe by rewriting it as follows %\teoadd{This equation is really small, we should fix it}
 \begin{eqnarray*}
		\lambda_{\mathit{crit}} &=& -(\lambda_2+1+\tilde{\Lambda})-(\Lambda+\tilde{\Lambda})+1\\
        & &+2\sqrt{(\Lambda+\tilde{\Lambda})(1+\lambda_2+\tilde{\Lambda})}\\
		&=& -\left[\sqrt{\lambda_2+\Lambda+\tilde{\Lambda}}-\sqrt{\Lambda+\tilde{\Lambda}}\right]^2+1 <1\, .
	\end{eqnarray*}

	\item If $\lambda_{\mathit{crit}}< \tilde{\lambda}_1<1$ and $1+\lambda_2>{\Lambda}$, then $\rho^*_\pm$ are real but negative, so they are not physically acceptable.
	\item If $\lambda_{\mathit{crit}}< \tilde{\lambda}_1<1$ and $1+\lambda_2<{\Lambda}$, then $\rho^*_\pm$ are real and positive, so they are both physically acceptable, moreover $\rho^*_{-}$ is unstable while $\rho^*_{+}$ is stable.
	\item If $\tilde{\lambda}_1>1$, $\rho^*_\pm$ are real but $\rho^*_{-}$ is negative, thus the only physically acceptable solution is $\rho^*_{+}$ which is stable.
\end{itemize}	  
By considering the behavior of the solutions $\rho^*_{\pm}$ for large $\tilde{\lambda}_1$, we can prove that $\rho^*_{+}\rightarrow 1$. In Table~\ref{tab:equilibria} we summarize the different asymptotic solutions of Eq.~\eqref{eq:edo} and their stability as a function of the involved parameters. 

The above analysis allows us to conclude that when $\Lambda > 1 + \lambda_2$, the prevalence of infected individuals has a discontinuous phase transition at $\tilde{\lambda}_1=\lambda_{\mathit{crit}}$ and it is bistable for $\lambda_{\mathit{crit}}< \tilde{\lambda}_1<1$. On the other hand when $\Lambda < 1 + \lambda_2$, the transition from the disease-free state to the endemic state is instead continuous.

\renewcommand\arraystretch{1.5}
\begin{table}[ht]
	\centering
	\caption{Number and stability of stationary density of infected people depending on the parameters $\tilde{\lambda}_1$ and $\Lambda$.}
	\resizebox{0.9\linewidth}{!}{
	\begin{tabular}[t]{l|c|c|c}
		%\hline
		$\tilde{\lambda}_1 < \lambda_{\mathit{crit}}$ & $ \lambda_{\mathit{crit}} < \tilde{\lambda}_1<1$ & $\lambda_{\mathit{crit}} < \tilde{\lambda}_1<1$ & $\tilde{\lambda}_1 \geq 1$ \\
		& $ \Lambda < 1 + \lambda_2$ & $\Lambda > 1 + \lambda_2$ & \\
		\hline
		$\rho^*_0=0$ (stable) & $\rho^*_0=0$ (stable) & $\rho^*_0=0$ and $\rho^*_{+}$ (stable)& $\rho^*_{+}$ (stable) \\
		&  & $\rho^*_{-}$ (unstable) & $\rho^*_0$ (unstable) \\
		\hline
		%\hline
	\end{tabular}
}
	\label{tab:equilibria}
\end{table}%

\subsection{Impact of $\eta$ on bistability}
\label{ssec:impactofeta}

\begin{figure}[t!]
	\centering
		\includegraphics[width=\columnwidth]{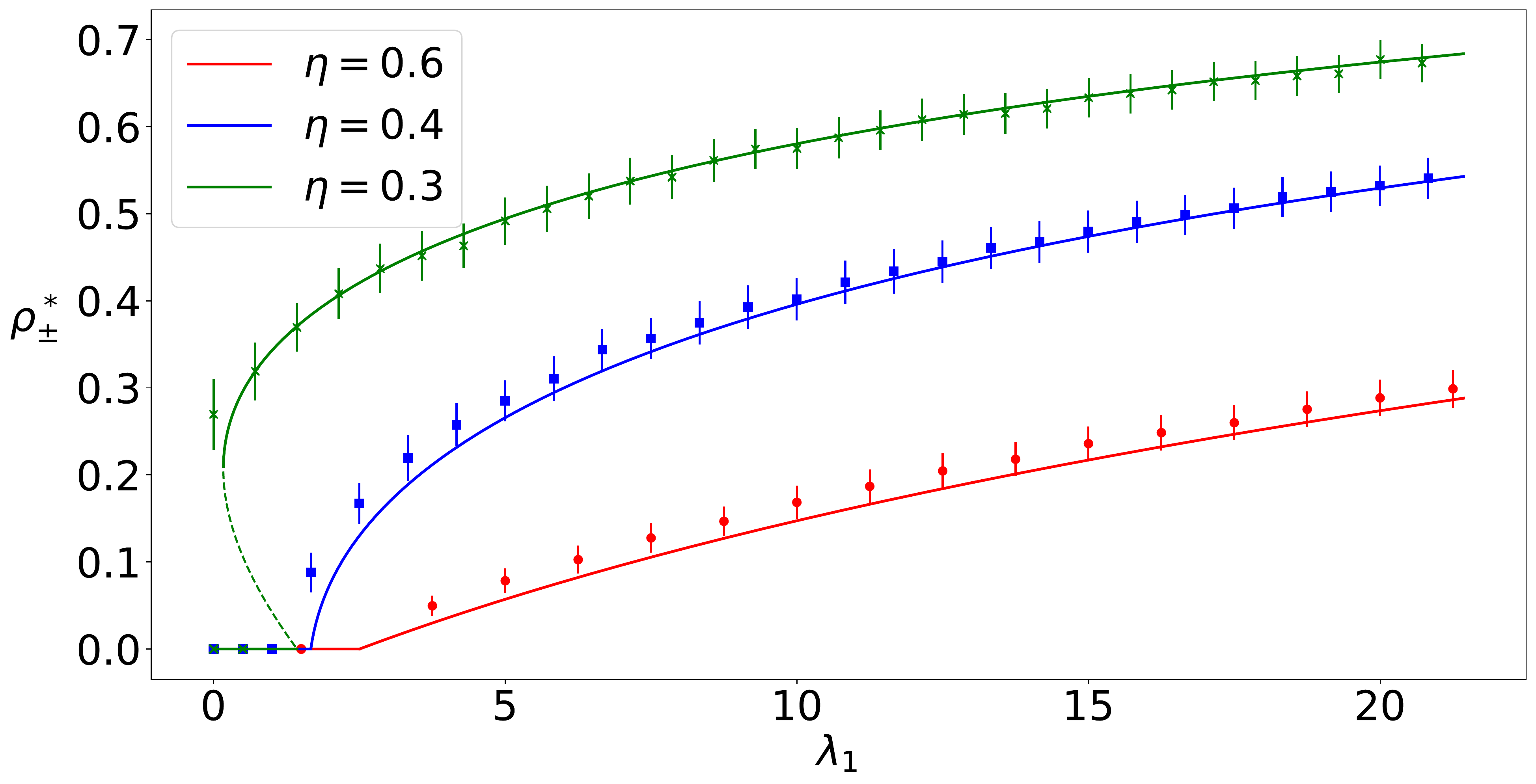}
		\caption{Prevalence $\rho^*$ as a function of $\lambda_1$ for several values of $\eta$ and $\mu = 0.07$, $\gamma_1 = 0.3$, $\gamma_2 = 0.2$, $\gamma_3=0$, $\gamma_4=0.2$, $\gamma_5=0$, $\beta_2 = 0.04$, $\langle k \rangle = 60$ and $\langle k_{\Delta} \rangle = 10$. Stochastic simulations (symbols) obtained with an implementation of the Gillespie algorithm are shown together with the mean-field prediction. Each symbol is the average over several independent realizations; the standard deviation is also reported (vertical bars).}
	\label{fig:rhostar_1}
\end{figure}

Results reported in Fig.~\ref{fig:rhostar_1} show that the prevalence $\rho^*$ increases with $\lambda_1=\frac{\beta_1 \langle k \rangle}{\mu}$ and as previously stated $\rho^*\rightarrow 1$ for unbounded values of $\lambda_1$. The mean-field predictions (curves) are compared with stochastic simulations (points) obtained by using the Gillespie algorithm; we can observe the very good agreement between the mean-field findings and the numerical ones, confirming thus the validity of the used assumptions. We refer to Appendix~\ref{app:stochastic} for a description of the algorithm and more details about the numerical simulations. 

The results shown in the figure allow us to appreciate the substantial impact that distrust has on the social contagion. First, we observe that for a given $\lambda_1$, which is above the epidemic threshold, $\rho^*$ decreases as $\eta$ increases. Hence, the larger is the distrust in the network the lower is the fraction of people that are infected at equilibrium. Even more interestingly, the presence of trust and distrust relationships conditions the nature of the transition from the disease-free state to the endemic one, and so the existence of a bistable regime. In particular, we observe that increasing the amount of distrust in the network makes the bistability region to shrink until it vanishes, with the transition at $\lambda_1 = 1/(1-\eta)$ becoming continuous. Mathematically, this means that by tuning the value of $\eta$ we can shift from a case where $\Lambda > 1 + \lambda_2$, i.e., the bistability region exists, to one where $\Lambda < 1 + \lambda_2$, i.e., a bistable regime is not allowed.

\subsection{Non-monotonic behavior of the prevalence %with respect to the degree connectivity $\langle k \rangle$
}
\label{ssec:nonmonot}
\begin{figure}[htb!]
	\centering
		\includegraphics[width=\columnwidth]{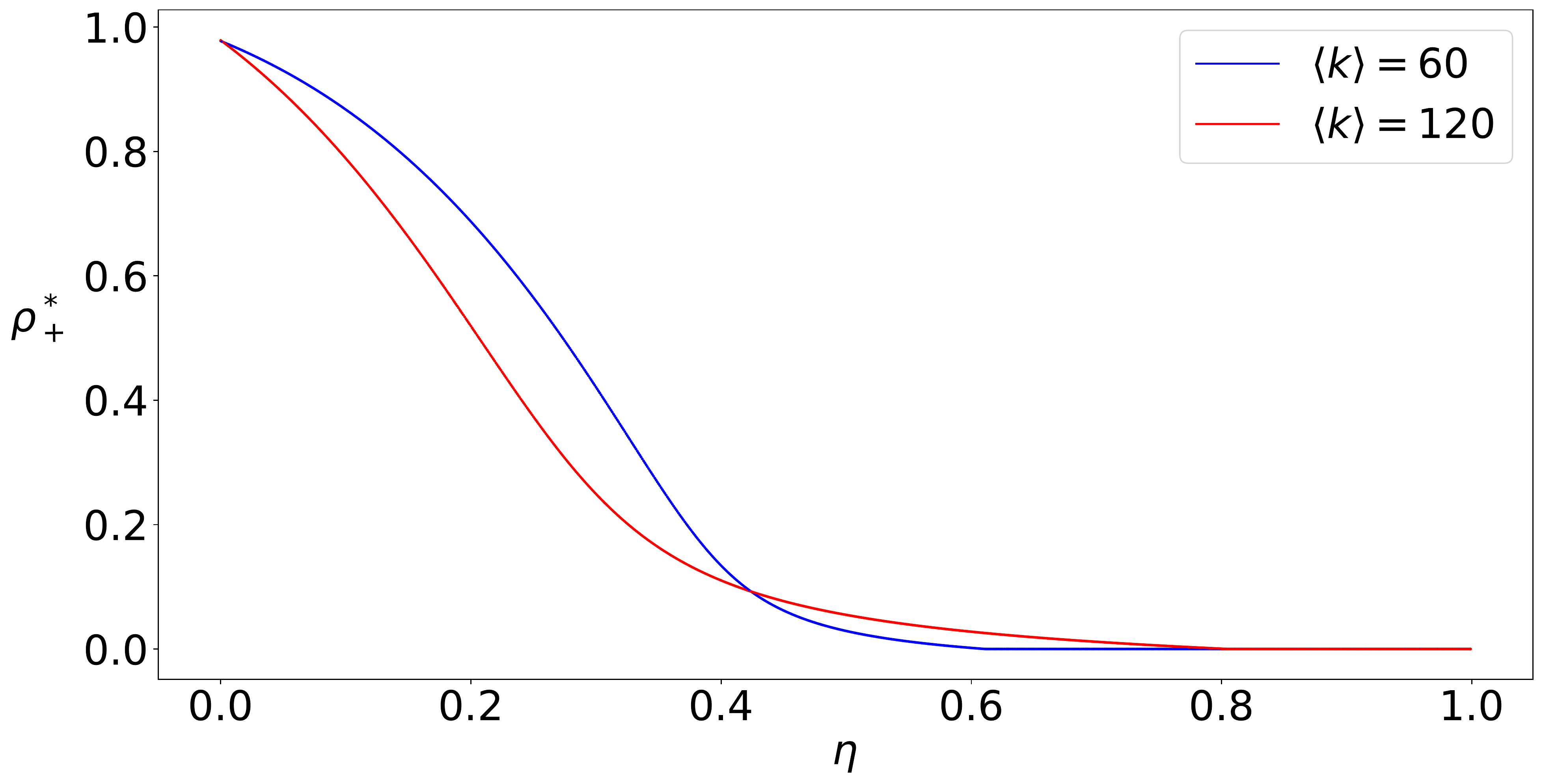}
		\caption{Prevalence as a function of $\eta$, for various values of $\langle k \rangle$, $\beta_1 = 0.003$ and remaining parameters identical to those of Fig.~\ref{fig:rhostar_1}.}
	\label{fig:rhostar_2}
\end{figure}

Let us now investigate how the infection depends on the average degree $\langle k \rangle$. In Fig.~\ref{fig:rhostar_2}, we show the prevalence $\rho^*_{+}$ as a function of the distrust $\eta$ for two particular values of the average node degree $\langle k \rangle$. The other parameters are identical to those used in Fig.~\ref{fig:rhostar_1}, but $\beta_1=0.003$. Each curve corresponds to a fixed different value of $\lambda_1$. Let us  consider the blue and red curves corresponding respectively to $\langle k \rangle = 60$ and $\langle k \rangle = 120$. The associated values of $\lambda_1$ (resp. $\sim 2.57$ and $\sim 5.14$) are large enough so that the system is never bistable regardless of $\eta$ (see Fig.~\ref{fig:rhostar_1}). 

As we have already seen, the lower the fraction of distrust and the larger the infection. We observe that, while the prevalence is monotonic and continuous with respect to $\eta$, it is non-monotonic with respect to the average connectivity $\langle k \rangle$. Indeed for low enough values of $\eta$ increasing the connectivity of the network reduces the prevalence while the opposite phenomenon takes place for large values of the distrust. Geometrically, this translates into the crossing of both curves at a particular value $\eta_{\text{cross}}$ which can be analytically computed (see Appendix~\ref{app:eta_cross}).

This behavior apparently looks counter-intuitive. For a given value of $\eta$, the only parameters changing as a function of $\langle k \rangle$ are $\lambda_1$ and $\lambda_2$. Therefore, the magnitude of the prevalence with respect to the average degree will depends on the relative strength between these two parameters, which are associated to pairwise infections and recoveries, respectively. 

\section{The social balance theory}
\label{sec:sbt}
The model presented in the previous section is based on the assumption that agents  
maintain the same pairwise trust/distrust relations when they are involved in interactions in group of more than two. 
This assumption implies 
that the distribution of signed triangles depends on the fraction of negative links in the network, $\eta$, and not on a peculiar combination of ``$+$'' and ``$-$''. However social balance theory affirms that in a social network, the fraction of balanced triangles, i.e., triadic relations with an odd number of positive links, is larger than the one we would obtain if signs were randomly distributed~\cite{heider1946attitudes,harary1953notion,szell2010multirelational}. 
In this section we extend the previous model in order to investigate the role of balanced and unbalanced full triangles on social contagion. This can be achieved by introducing new parameters, $p_i\geq 0$, where $i=0,\dots,3$, refers to the number of positive links in the triangles, to express the excess or the deficiency of peculiar signed $2$-simplices with respect to the case where these inherit the signs of the links. This allows us to write
\begin{eqnarray}
\label{eq:nubalth}
\tau_{0}&=&p_0\eta^3\langle k_\Delta\rangle\, , \tau_{1}=3p_1\eta^2(1-\eta)\langle k_\Delta\rangle\notag\\
\tau_{2}&=&3p_2\eta(1-\eta)^2\langle k_\Delta\rangle\text{ and }\tau_{3}=p_3(1-\eta)^3\langle k_\Delta\rangle\, .
\end{eqnarray}
The model presented in the previous section, i.e., where all the signs were randomly attributed to the edges, can be recovered with the choice $p_i = 1$ for all $i$. If there were only balanced triangles, then $p_0 =p_2 = 0$, while $p_1 =  p_3=0$ denotes the complete absence of balanced triangles. Let us note that the parameters $p_i$ are not independent from each other as we must have $\sum_i \tau_{i} = \langle k_\Delta \rangle$, hence
\begin{equation}
p_0\eta^3+3p_1\eta^2(1-\eta)+3p_2\eta(1-\eta)^2+p_3(1-\eta)^3=1\, .
\label{eq:sum}
\end{equation}

Let us observe that the solutions of Eq.~\eqref{eq:main} are still given by $\rho^*_0=0$ and $\rho^*_\pm$ obtained by~\eqref{rhopm}, of course with new parameters $\Lambda_i$, $i=1,\dots,5$ depending on the new definitions of $\tau_{k}$, $k=0,\dots,3$ given by~\eqref{eq:nubalth}. The same analysis above performed concerning the existence and stability of the stationary densities can be straightforwardly adapted to the present case.

\begin{figure}[t!]
	\centering
	\begin{tabular}{cccc}
		\includegraphics[width=\columnwidth]{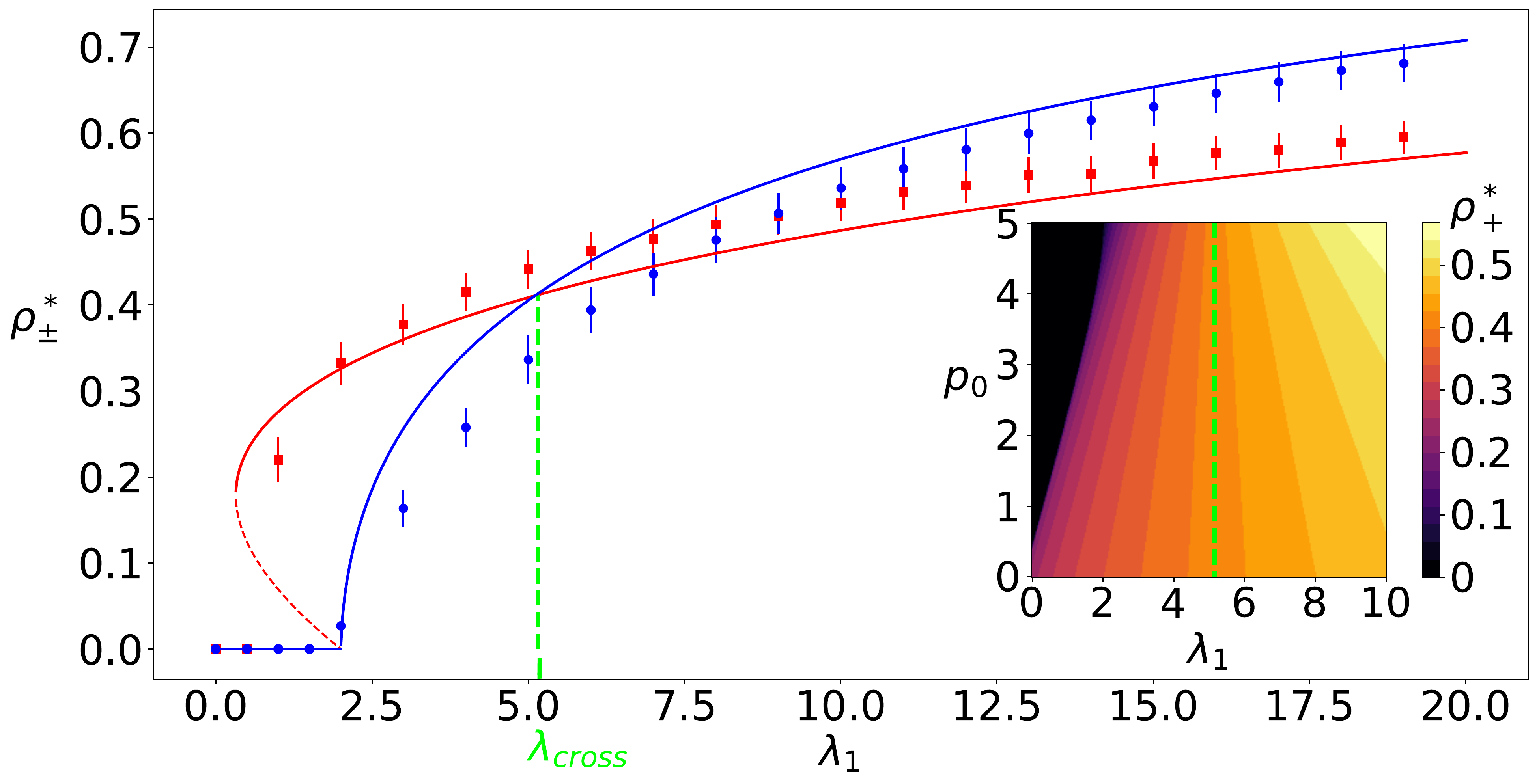}
	\end{tabular}	
	\caption{Prevalence $\rho^*$ as a function of $\lambda_1$ for $p_0=1$ (red curve) and $p_0=5$ (blue curve) and fixed parameters $p_2=1$, $\eta = 0.5$, $\langle k \rangle = 60$, $\langle k_\Delta \rangle = 10$, $\mu=0.05$, $\beta_2 = 0.007$, $\gamma_1 = 0.35$, $\gamma_2 = 0.2$, $\gamma_3 = 0$, $\gamma_4 = 0.5$ and  $\gamma_5 = 0$. The parameters $p_1=p_3$ are deduced from Eq.~\eqref{eq:sum} {and take the value $1$ for $p_0=1$ (red curve) and $0$ for $p_0=5$ (blue curve)}. Inset: Phase diagram of the infection $\rho^*_+$ as a function of the parameters $p_0$ and $\lambda_1$ for the same set of parameters.}
	\label{fig:influence_pi}
\end{figure}

In the following, we examine how the distribution of distrust relations influences the contagion outcome. As a first analysis, we set the bias parameters relative to balanced triangles to be equal, i.e., $p_1 = p_3$, and we investigate the impact on the contagion of the parameters relative to unbalanced triangles, i.e., $p_0$ and $p_2$. In Fig.~\ref{fig:influence_pi} we show the prevalence curves for two values of $p_0$, i.e., $p_0=1$ (red curve) and $p_0=5$ (blue curve), while fixing $p_2 = 1$, the values of the other parameters are reported in the caption.
%$\eta = 0.5$, $\langle k \rangle = 60$, $\langle k_\Delta \rangle = 10$, $\mu=0.05$, $\beta_2 = 0.007$, $\gamma_1 = 0.35$, $\gamma_2 = 0.2$, $\gamma_3 = 0$, $\gamma_4 = 0.5$, $\gamma_5 = 0$. 
Notice that the value of $p_1=p_3$ is automatically adapted as to satisfy Eq.~\eqref{eq:sum}. Note also that $\eta$ is fixed, meaning that the average number of negative links is constant. In the same Figure we compare the mean-field solution (curves) with the results of stochastic simulations performed using the Gillespie algorithm (points); the agreement is again satisfactory, the discrepancy being due to finite size effects that enhance correlations, neglected in the construction of the mean-field model (see Appendix~\ref{app:stochastic} for more details). 

We can observe two opposite behaviors. On the one hand, for small values of $\lambda_1$, the prevalence $\rho^*$ increases when reducing $p_0$, i.e., decreasing the number of unbalanced triangles (the red curve is above the blue one). On the other hand, for large values of $\lambda_1$, we find on the contrary that $\rho^*$ increases as we increase $p_0$ (the red curve is below the blue one). This implies the existence of a value $\lambda_{\mathrm{cross}}$ at which both curves have the same prevalence. The prevalence at the particular value $\lambda_{\text{cross}}$ can be computed explicitly (see Appendix \ref{app:lambda_cross}) and is given, by assuming $\eta=1/2$, $p_1=p_3$ and $p_2=1$ by:
\begin{equation}
    \rho^*(\lambda_{\text{cross}}) = \frac{\gamma_1}{\gamma_1 + \gamma_4 - 4\gamma_5},
\end{equation}
provided the parameters $\gamma_4 > 4 \gamma_5$ to ensure the prevalence to lie into $(0,1)$. For the parameters used in Fig.~\ref{fig:influence_pi}, we obtain $\rho^*(\lambda_{\text{cross}})=7/17\sim 0.41$. 

To further investigate the crossing behavior, we construct the phase diagram of $\rho^*$ as a function of $p_0$ and $\lambda_1$. This is shown in the inset Fig.~\ref{fig:influence_pi}. Note that we binned the values of $\rho^*$ so to better visualise the crossing. For large values of $\lambda_1$ (right of the dashed green line), the prevalence increases as we increase $p_0$, while the opposite holds for small values of $\lambda_1$ (left of the dashed green line). Note that the black region corresponds to the parameter set for which the disease-free state is the only stable equilibrium. 

These results highlight how the effect of balanced and unbalanced triangles on the prevalence highly depends on the link infectivity of the process. As already stated, real social systems are more likely composed by balanced configurations, i.e., they are associated to low values of $p_0$ and $p_2$. So in our analysis, we expect a contagion spreading on them to show a behavior more similar to that of the red curve. Hence, we would expect real social systems to have larger (resp. smaller) prevalence compared to random sign configurations when the infectivity $\lambda_1$ is small (resp. large). This counter-intuitive behavior is due to the high nonlinearity of the contagion process, as infection and recovery mechanisms are associated to both balanced and unbalanced triangles (see Fig.~\ref{fig:case1} and \ref{fig:case2}).

Finally, we find that by tuning the bias parameter $p_0$ we can change the nature of the transition from the disease-free state to the endemic one. Indeed, we can observe that the system becomes bistable for low enough values of $p_0$ (the red curve shows bistability, while the blue curve does not). 

\begin{figure}[htb!]
	\centering
	\begin{tabular}{cccc}
		\includegraphics[width=0.42\columnwidth]{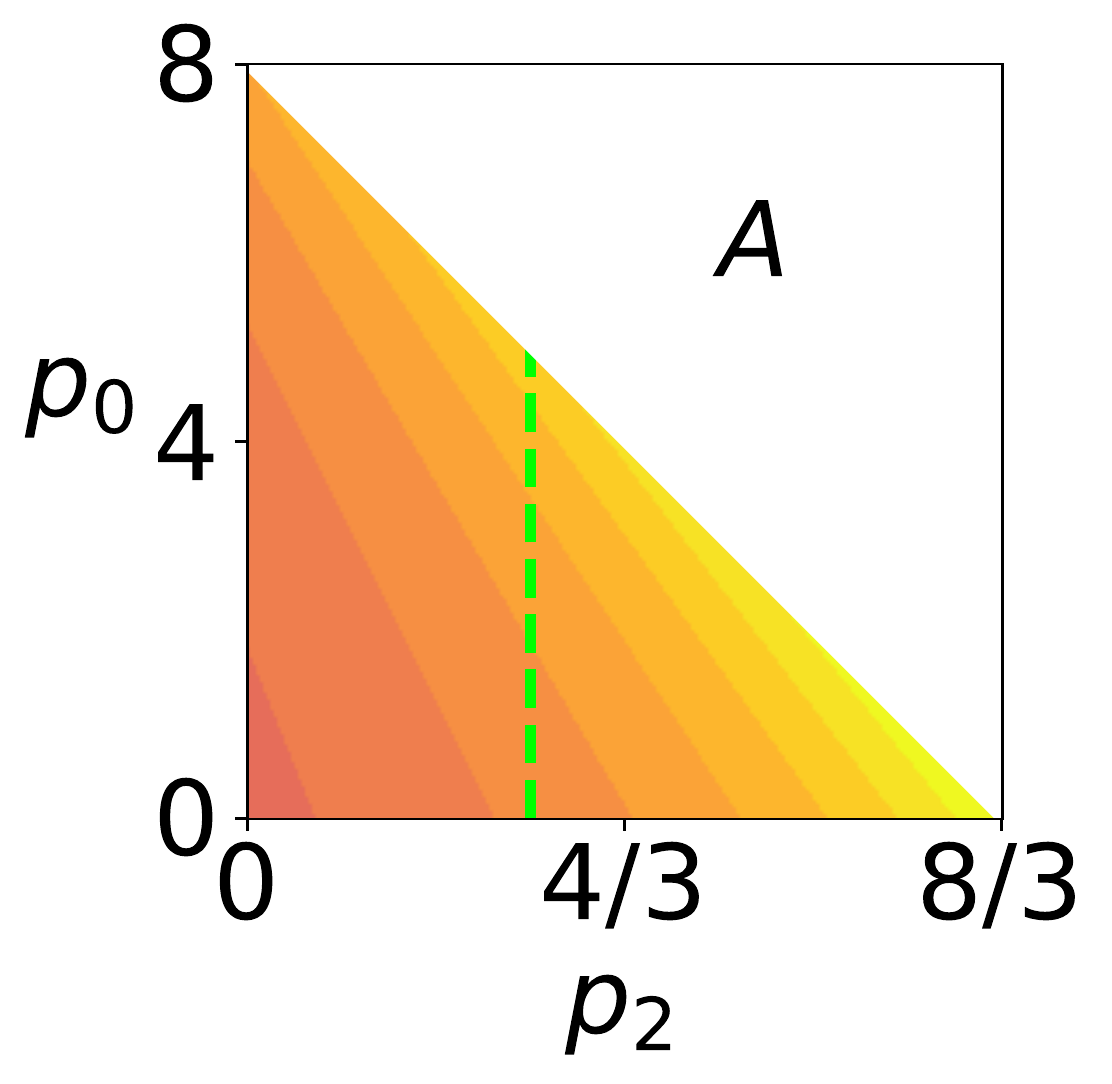}
		\includegraphics[width=0.5\columnwidth]{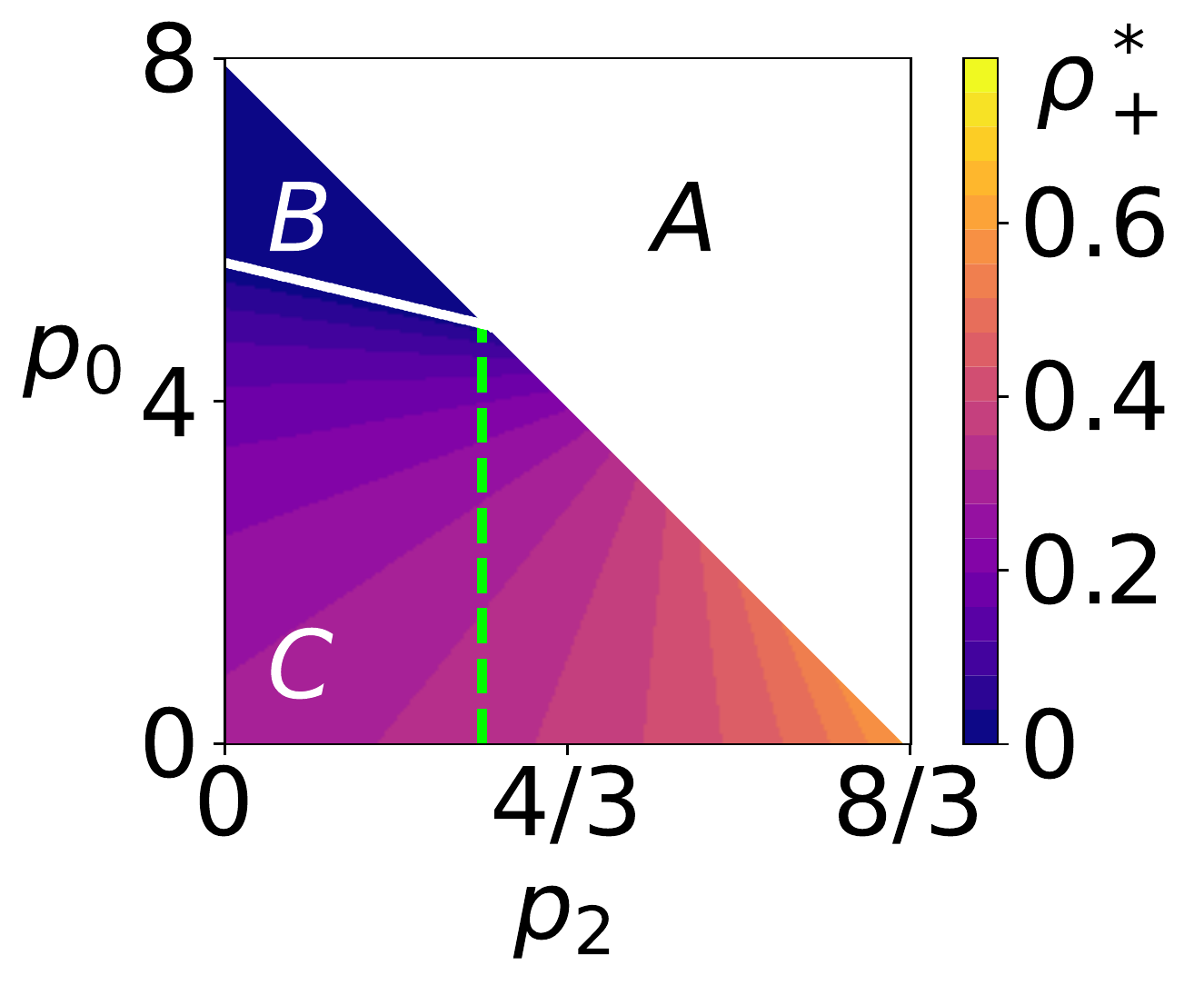}
	\end{tabular}	
	\caption{Phase diagram of the infection $\rho^*$ as a function of the parameters $p_0$ and $p_2$ for $\lambda_1=20$ (left) and $\lambda_1=2$ (right) and the remaining parameters identical to those of Fig. \ref{fig:influence_pi}. The vertical dashed curve (in green) corresponds to $p_2=1$.}
	\label{fig:influence_pi_2}
\end{figure}

We now allow $p_2$ to vary, thus studying the behavior of the system as a function of both $p_0$ and $p_2$. Still, we assume that $p_1=p_3$, with their value determined by Eq.~\eqref{eq:sum}. In Fig.~\ref{fig:influence_pi_2}, we show the phase diagram of $\rho^*$ as function of $p_0$ and $p_2$ for $\lambda_1 = 20$ (left) and $\lambda_1 = 2$ (right). In both panels, the white region denoted by $A$ is such that the values of $p_0$ and $p_2$ would imply $p_1=p_3<0$ which is not physically acceptable. In particular, the delimiting line of region $A$ is given by $p_0\eta^3+3p_2\eta (1-\eta)^2=1$, and it corresponds to the case where no balanced triangles are present, i.e., $p_1 = p_3 = 0$. Coherently with the previous analysis, when $\lambda_1$ is large enough (left panel), the infection increases with $p_0$ and $p_2$. Note that the vertical dashed line represents $p_2=1$, corresponding to the case previously studied. For $\lambda_1=2$, the behavior of the system with respect to $p_0$ and $p_2$ is more complex. Indeed, for small values of $p_2$, the prevalence decreases as $p_0$ is increases, while the opposite is true for large values of $p_2$. 

For this latter case, we distinguish two additional zones. In zone $B$, there is no bistability, while the latter is present in zone $C$. The separation line between the two is found for $\Lambda = 1+\lambda_2$, and takes, for the above values of the parameters, the expression, $p_0 = -\frac{5}{7}p_2 + \frac{984}{175}$.

To conclude the analysis, we further investigate the impact of the particular configurations of balanced and unbalanced triangles present in the network. In particular, we first fix the values of $p_1$ and $p_3$, namely the bias parameters for the balanced triangles, and vary $p_0$ and $p_2$, so the bias parameters for the unbalanced configurations, then we do the opposite, fixing $p_0$ and $p_2$, and varying $p_1$ and $p_3$.

\begin{figure}[htb!]
	\centering
		\includegraphics[width=\columnwidth]{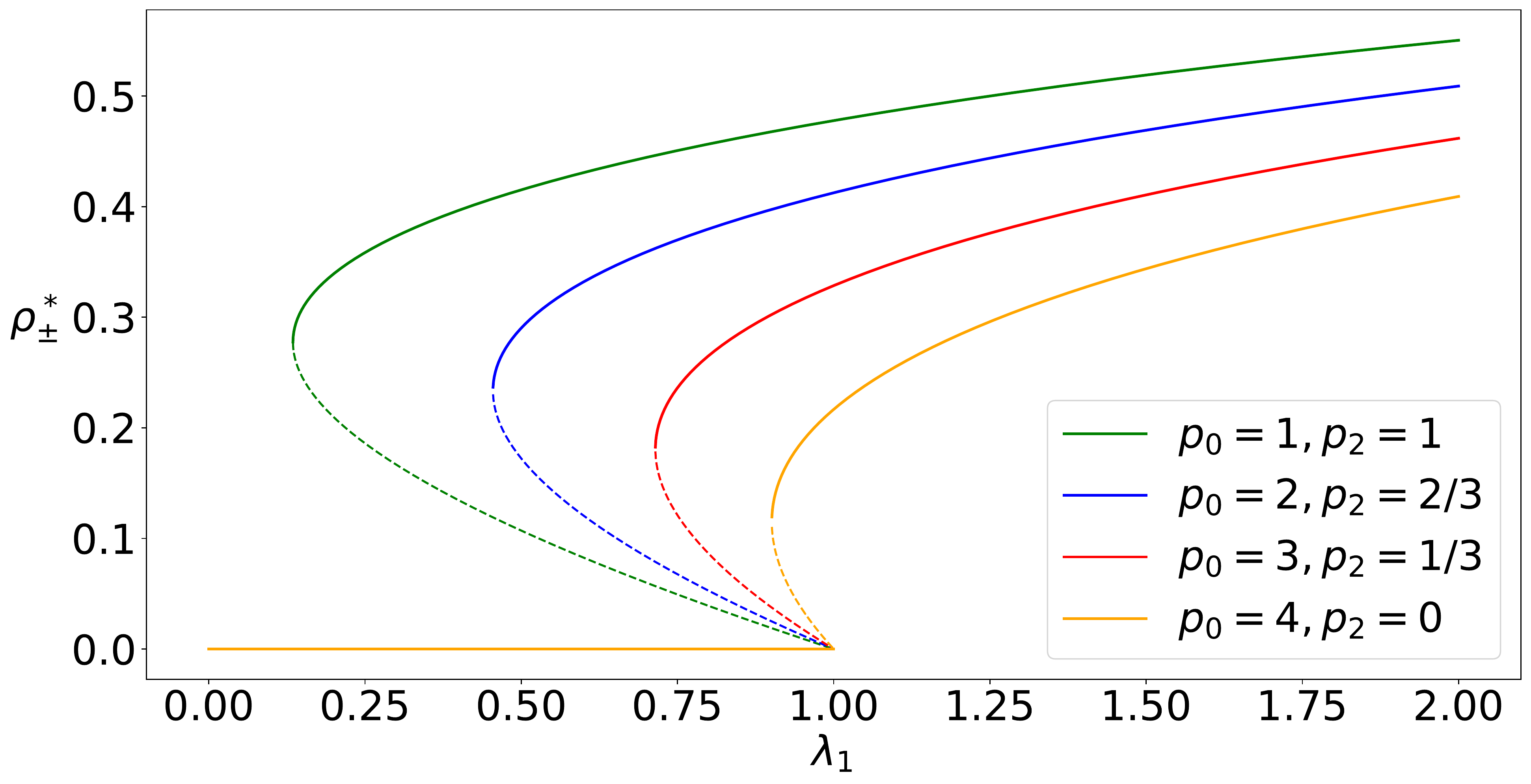}
		\includegraphics[width=\columnwidth]{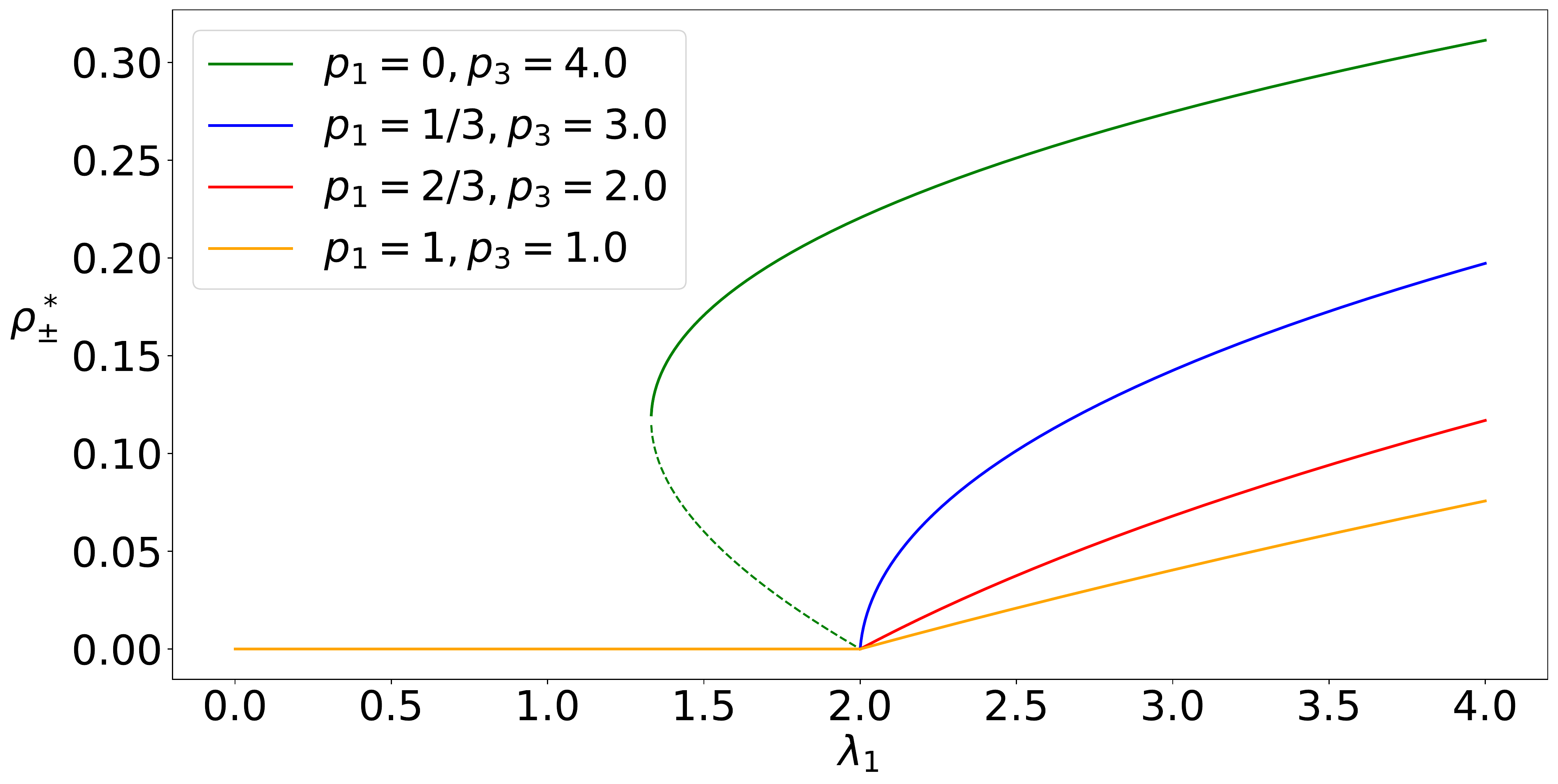}
	\caption{{Top: Fraction $\rho^*$ of infected nodes as a function of $\lambda_1$ for distinct values of $p_0$ and $p_1=p_3=1$, $\eta = 0.5$, $\mu = 0.05$, $\beta_2 = 0.01$, $\gamma_1 = 0.15$, $\gamma_2 = 0.1$, $\gamma_3 = 0$, $\gamma_4 = 0.05$, $\gamma_5 = 0$, $\langle k \rangle = 30$ and $\langle k_{\Delta} \rangle = 15$. Bottom: Fraction $\rho^*$ of infected nodes as a function of $\lambda_1$ for distinct values of $p_1$ and $p_0=p_2=1$, $\eta = 0.5$, $\mu = 0.05$, $\beta_2 = 0.06$, $\gamma_1 = 0.15$, $\gamma_2 = 0.05$, $\gamma_3 = 0$, $\gamma_4 = 0.05$, $\gamma_5 = 0$, $\langle k \rangle = 30$ and $\langle k_{\Delta} \rangle = 15$. } }
	\label{fig:perfect_un_balance}
\end{figure}

In the top panel of Fig.~\ref{fig:perfect_un_balance} we report the fraction of infected nodes as a function of $p_0$ and $p_2$ for fixed $p_1 = p_3 = 1$; let us remember again that because of the constraint~\eqref{eq:sum} $p_0$ and $p_2$ are not independent each other. We can observe that for a fixed $\lambda_1$, the larger $p_2$, i.e., the more triangles with two positive links, the larger the final infection state. Let us emphasize that $p_0$, i.e., the fraction of triangles with no positive links, impacts the system outcome in the opposite direction: the larger the value of the parameter, the smaller the fraction of infected. This comes from the fact that $p_2$ is associated to the three-body infection process having rates $\gamma_2$ and $\gamma_3$ (see Fig.~\ref{fig:case1}), while $p_0$ is related instead to recoveries occurring at rate $\gamma_5$ (see Fig.~\ref{fig:case2}). Let us also observe that by increasing $p_2$, and thus by decreasing $p_0$, $\lambda_{\mathrm{crit}}$ decreases and thus the system is more prone to exhibit a bistable regime. Moreover, for a given value of $\lambda_1$ in the bistability region, we find that the critical amount of infected people needed to converge to the endemic state decreases as $p_2$ increases. Stated differently, by increasing the number of unbalanced triangles with two positive links, the system can ``easily'' switch to an endemic state.

In the bottom panel of Fig.~\ref{fig:perfect_un_balance} we show the results corresponding to a variation of $p_1$, and thus $p_3$, for fixed $p_0=p_2=1$. Again the two parameters have opposite impact: by increasing $p_3$ (resp. $p_1$), the infection $\rho^*$ increases (resp. decreases). Similarly to the previous case, $p_3$ is associated to three-body infections occurring at rate $\gamma_1$, while $p_1$ is linked to recoveries occurring at rate $\gamma_4$. Coherently, when we increase the value of $p_3$ a bistable regime can emerge.
%the critical value $\lambda_1$ is larger and the tipping point $\lambda_\mathrm{crit}$ is smaller; the system hence often shows an endemic free behavior, but can be ``easily'' lost if the initial fraction of infected people is sufficiently large.

Overall these results suggest that the bias parameters $p_i$ strongly impact the outcome of the contagion process, both in terms of its final prevalence and of its dynamics, i.e., continuous and discontinuous transitions, in a nontrivial way. Also, our model highlights that not only the relative proportion of balanced and unbalanced triangles determines the contagion, but also the particular configurations within these two classes can have a crucial influence on the system.

\section{Conclusion}
\label{sec:conclusion}
In this paper we have introduced a model of complex contagion on signed higher-order networks, which has allowed us to investigate, both numerically and  analytically, the combined effect of group interactions and distrust relations on the dynamics of social contagion.   
In our model, a social system with group interactions 
is represented as a signed simplicial complex, in which the edges are attributed either a ``$+$'' or a ``$-$'' sign to respectively denote trust or  distrust between agents. 
In our model we have two compartments, i.e. nodes can either be in the infected state (if they have adopted an opinion or a social norm) or the susceptible state. 
The existence of signed relations  allows for additional infection and recovery channels, which are different from those in standard model for the spreading of a disease. Namely 
an infected agent can become susceptible because of a signed three-body interaction. 
Two variants of the model were considered: in the first one, edge signs were attributed randomly and maintained the same during a group interaction, while in the second one the distribution of signs in the triadic groups was biased so as to account for social balance theory. 
With the first variant of the model, we have highlighted how trust and distrust relations affect the transition of a system from a disease-free state to an endemic one. 
Specifically, we have shown that increasing distrust can change the nature of the transition from discontinuous to continuous, by making the bistability region associated to the first-order transition vanish.
Moreover, we have characterized the nontrivial interplay existing between the fraction of distrust relations and the connectivity of the social network in shaping the collective dynamics of the system.
With the second variant of the model, we have analyzed how the precise patterns of trust/distrust relations impact complex contagion in higher-order networks. 
In particular, we have highlighted how the contagion is determined not only by the relative proportion of balanced and unbalanced triangles, but also by which configuration within these two classes is more frequent.
Various extensions of the present work are possible.
% 0) real-world network - same model but for real
% 1) beyond the 3 nodes - same mechanisms, more nodes
% 2) temporal network - same mechanisms, with time
% 3) different models - different mechanisms 
% 4) No Mean Field - different topology
First of all, it would be interesting to investigate how trust and distrust influences the contagion process on real-world higher-order networks. Also, it would be useful to generalize the model by considering larger group structures, e.g., groups of size $4$, and the new infection and recovery channels related to those larger groups. A third research direction would be to allow the pattern of contacts to change over time or the signed relations and nodes states to co-evolve. Finally, it would be interesting to consider an heterogeneous mean-field approach to better investigate the impact of the structural properties of higher-order networks on contagion.
%For instance, we here assumed pattern of contacts and (dis)trust relations to be constant over time but it would be of interest to relax such assumptions by using for instance a temporal network structure or by allowing signed relations to evolve according to the nodes states.%

%\clearpage
\bibliographystyle{ieeetr}
\bibliography{biblio}

\newpage
\onecolumngrid

\appendix

\section{Stochastic simulations}
\label{app:stochastic}

In this {appendix}, we {explain the procedure employed to generate (synthetic) signed simplicial complexes. We also briefly recall the implementation of the Gillespie algorithm used to simulate the contagion process on top of simplicial complexes.} 

{To construct a simplicial complex made of $N$ nodes with an average number $\langle k \rangle$ of links incident to a node and an average number $\langle k_{\Delta} \rangle$ of triangles incident to a node, we used the procedure described in~\cite{iacopini2019simplicial}.} {Given a set of $N$ nodes, } we first consider every pair of nodes and connect them with a probability 
\begin{equation*}
p_1 = \frac{\langle k \rangle -2\langle k_\Delta \rangle}{(N-1)-2\langle k_\Delta \rangle}\, .
\end{equation*}
We then run over all the triplets of nodes and {promote each of them to a} $2$-simplex with probability
\begin{equation*}
p_2 = \frac{2\langle k_\Delta \rangle}{(N-1)(N-2)}\, .
\end{equation*}
{One can check that such probabilities lead to the desired average degree $\langle k \rangle$ and hyperdegree $\langle k_\Delta \rangle$.}
Once the simplicial complex is constructed we have to associate signs to links and triangles. In the first model, we consider successively all the links and attribute {them} a negative sign with a probability $\eta$ and a positive sign otherwise. Then because we assume the agents keep the same dyadic trust/distrust relations once in the three-body  interaction, we check all the $2$-simplices and we add the same signs of the three pairwise couples.

In the second model, we take into account social balance theory by using a biased distribution of signs at the level of the triangles. In particular, the pairwise interactions receive a sign with the same process as above, i.e.,  a ``$-$" sign with a probability $\eta$ and a ``$+$" sign otherwise. On the other hand the signs of each triangle are assigned with the probabilities given by Eq.~\eqref{eq:nubalth} for a given choice of $p_i$, $i=0,\dots, 3$, namely we attribute $i$ positive signs and thus $3-i$ negative signs with a probability $\frac{\tau_{i}}{\langle k_{\Delta} \rangle}$.

As already described in the main text, we performed the stochastic numerical simulations by using the Gillespie algorithm, that we here briefly describe; we refer the interested reader to~\cite{gillespie,CIANCI201466,CarlettiFilisetti2012} for more details and other applications of the method. At time $t_0=0$, we start with a chosen fraction of infected nodes uniformly randomly distributed in the simplicial complex. We then establish a list of all the possible reactions, i.e., interaction channels as given in Figs.~\ref{fig:case0},~\ref{fig:case1} and~\ref{fig:case2}, that can occur given the states of the nodes and the structure of the signed simplicial complex. The Gillespie method is able to determine the most probable reaction to occur and the most probable time {at which} it will {take place}. By denoting with $r_i$ the rate associated with the $i$-th reaction and by $a_0=\sum_{i=1}^M r_i$ the sum of all the rates, we generate two random numbers $r_1$ and $r_2$ from the uniform distribution in $(0,1)$. The next reaction that will occur is the $\mu$-th with $\mu$ such that:
\begin{equation*}
\sum_{j=1}^{\mu-1} r_{j} < r_2 a_0 \leq \sum_{j=1}^{\mu} r_j\, .
\end{equation*}
This reaction will occur at time $t_1=t_0+\tau=\tau$ with 
\begin{equation*}
\tau = \frac{1}{a_0} \ln (1/r_1)\, .
\end{equation*}
Note that the underlying assumption is that only one node changes its state during each reaction. After updating the state of this node, we repeat the process until the fraction of infected nodes stabilizes, i.e., fluctuates around the mean-field equilibrium taking into account the fluctuations.

\begin{figure}
	\centering
	\includegraphics[width = 0.49\columnwidth]{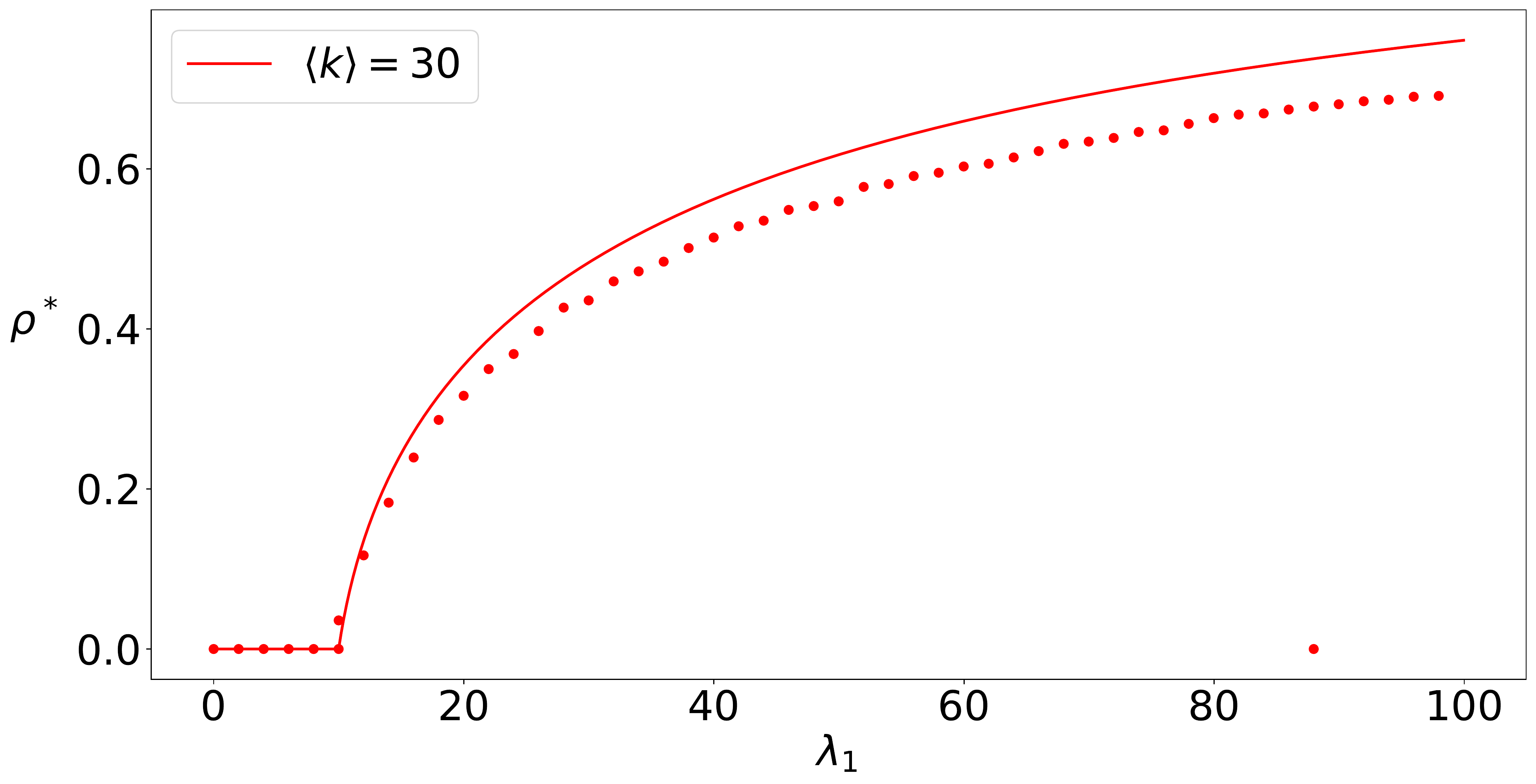}
    \includegraphics[width = 0.49\columnwidth]{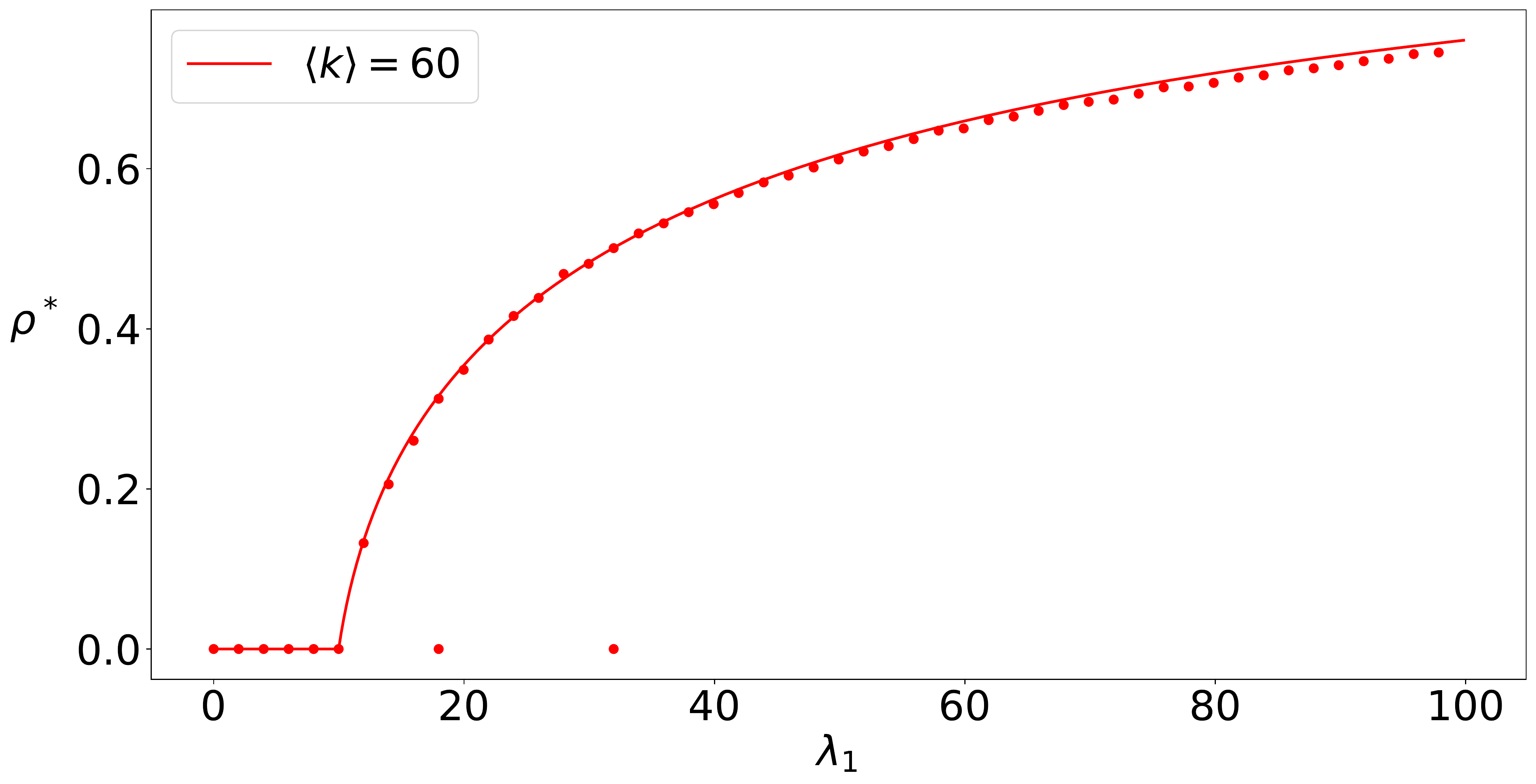}
	\caption{{Prevalence $\rho^*$ as a function of $\lambda_1$ in a simplex made by nodes $N = 1000$ and parameters $\mu = 0.05$, $\beta_2 = 0$, $\gamma_1 = 0.2$, $\gamma_2 = 0$, $\gamma_3 = 0$, $\gamma_4 = 0.1$, $\gamma_5 = 0$, $\langle k_{\Delta} \rangle = 15$, $p_1 = 1$, $p_2 = 1$, $p_3 = 1$, $p_4 = 1$, $\eta = 0.9$ with $\langle k \rangle = 30$ (left panel) and $\langle k \rangle = 60$ (right panel). {The mean-field prediction (continuous curve) is compared with the stochastic simulations (red points) performed by using the Gillespie algorithm.}}}
	\label{fig:Gillespie1}
\end{figure}

{Fig.~\ref{fig:Gillespie1} compares, for two distinct values of $\langle k \rangle$ but fixed remaining parameters, the stochastic simulations (points) with respect to the mean-field prediction (continuous curve). The fraction of infected nodes obtained with the Gillespie algorithm was averaged over the last $10$\% of the orbit (when the system is close to the stationary state).} To detect possible bistability, for each value of $\lambda_1$, we ran several independent stochastic simulations, with a distinct fraction of initially infected nodes, ranging in $(0,1)$, uniformly randomly distributed among the nodes of the simplicial complex. Each stochastic simulation corresponds to a newly generated simplicial complex satisfying the above constraints on $\langle k \rangle$, $\langle k_{\Delta} \rangle$ and the fraction of signed triangles. 

Let us observe that for $\langle k \rangle = 30$, the stochastic simulations {deviate from the mean-field predictions}. Nevertheless, increasing the network density, e.g., by taking $\langle k \rangle = 60$ reduces this discrepancy as can be seen in {the right panel} for the same value of $\eta$. In some few cases we observe a vanishing infection for large values of $\lambda_1$. This epidemic fade-out happens when initially the fraction of infected nodes is very small in which case the infection might die out due to the stochasticity of the system and to finite size population. Hence by increasing the number of nodes in the network, such behavior should disappear.

\clearpage
\section{Non-monotonicity {of the prevalence} with respect to the {average node degree}}
\label{app:eta_cross}
\begin{figure}[h]
	\includegraphics[scale=0.37]{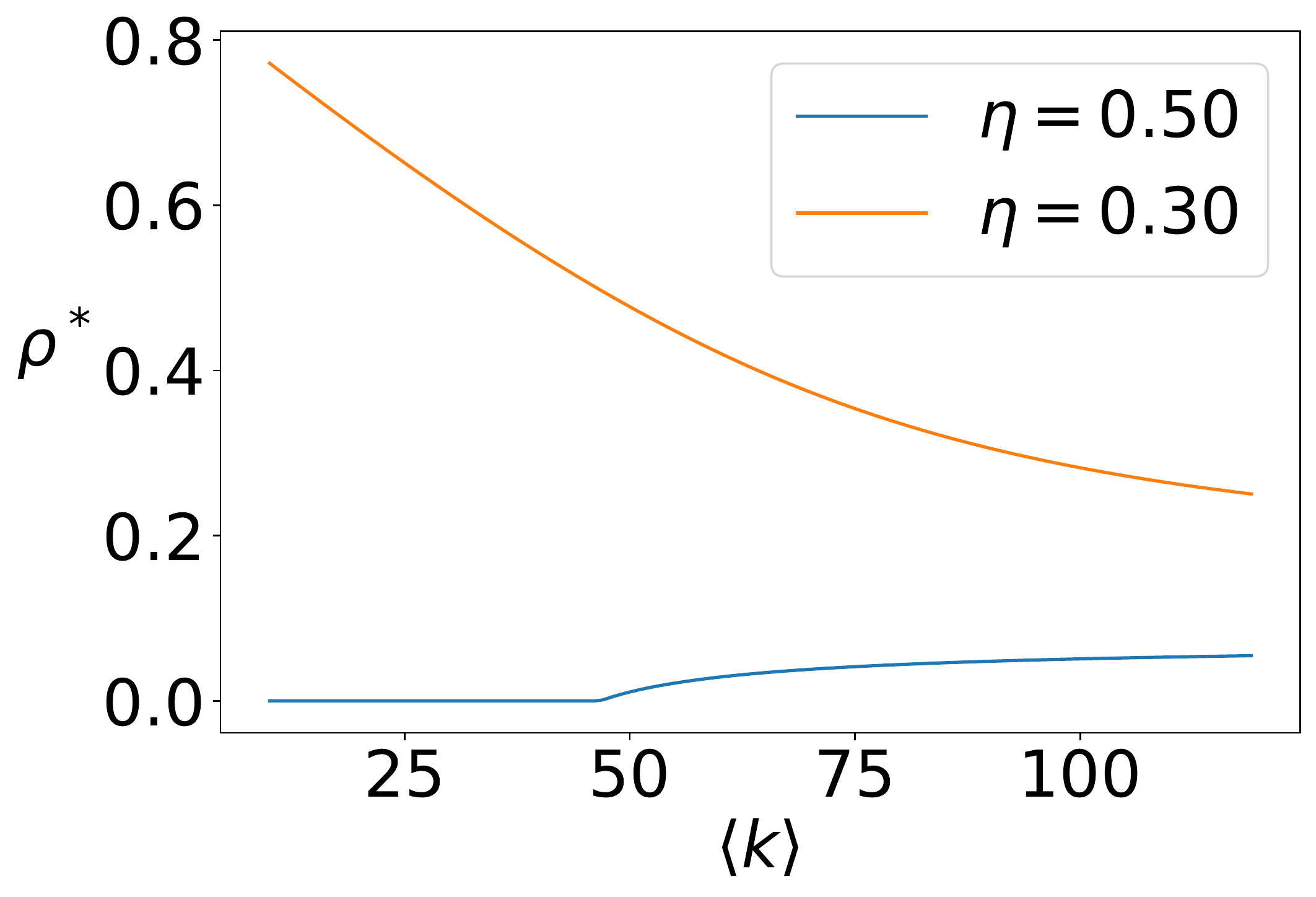} \hfill
	\includegraphics[scale=0.37]{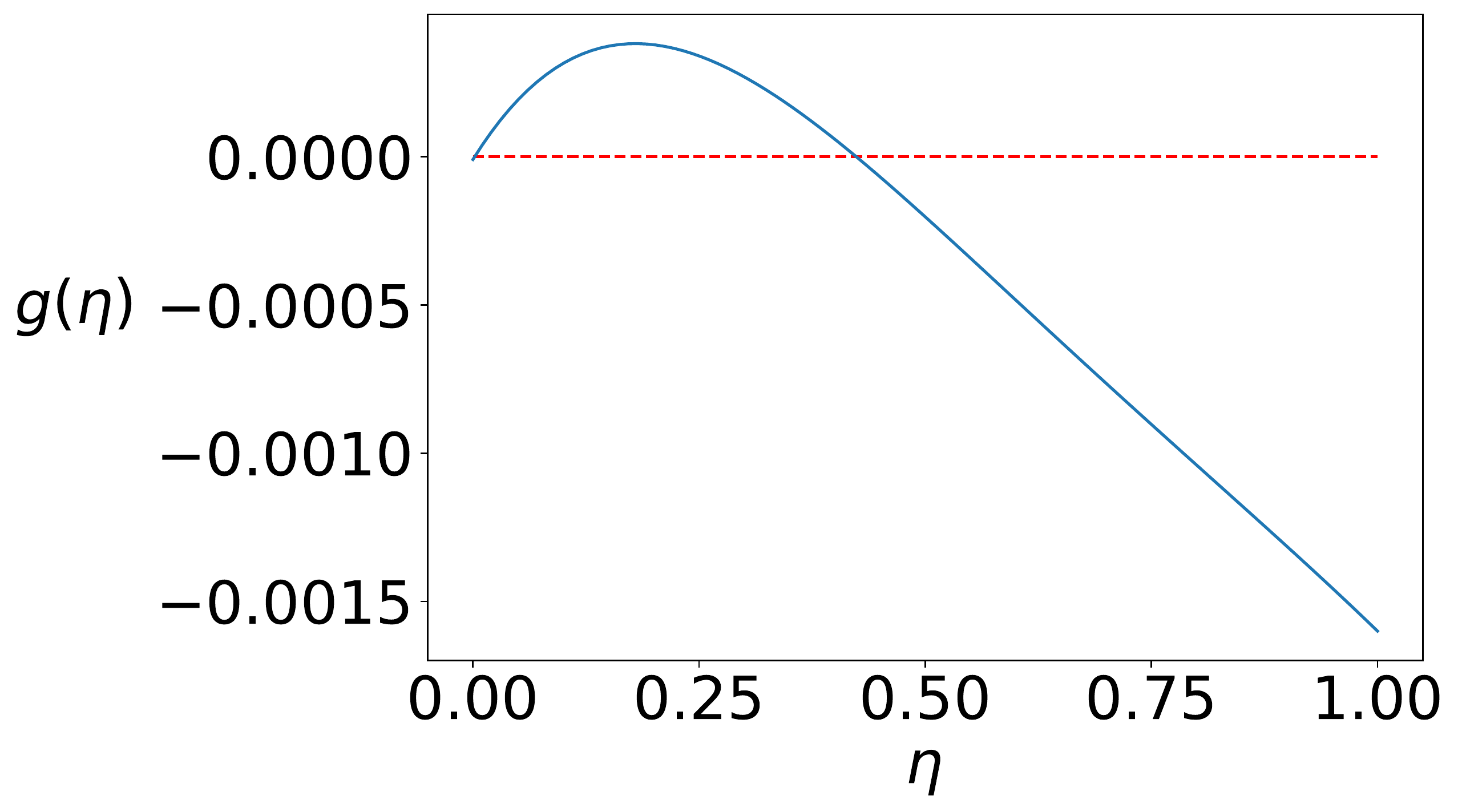}
	\caption{Left: Infection as a function of the average node degree $\langle k \rangle$. When $\eta < \eta_{\text{cross}}$ with $g(\eta_{\text{cross}}) = 0$, the infection decreases with $\langle k \rangle $ while it is the opposite when $\eta > \eta_{\text{cross}}$. The parameters are: $\beta_1 = 0.003, \beta_2 = 0.04, \gamma_1 = 0.3, \gamma_2 = 0.2$, $\gamma_3=0, \gamma_4=0.2, \gamma_5=0, \mu=0.07$ and $\langle k_\Delta \rangle =10$. Right: Plot of the function $g(\eta)$ for the aforementioned parameters.}
	\label{fig:Non_monotonicity}
\end{figure}

In this appendix, we analyze the non-monotonic behavior of the infection as a function of the average node degree $\langle k \rangle$ and the fraction of distrust $\eta$ in the network.
Let us recall that the stationary density is given by 
\begin{equation}
\rho^*_{\pm} = \frac{
	\Lambda - \lambda_1- \lambda_2 
	\pm \sqrt{ {(\Lambda- \lambda_1- \lambda_2)}^2
		-4(\Lambda + \tilde{\Lambda})(1-\lambda_1)}
}{
	2(\Lambda + \tilde{\Lambda})\, 
},
\end{equation}
see Eq.~\eqref{rhopm} in the main text. To highlight the non-monotonic behavior of the infection with respect to the average node degree $\langle k \rangle$, we compute the derivative of $\rho^*_+$ with respect to $\langle k \rangle$ and determine for which value of $\eta$ it vanishes. One finds:
\begin{equation}
    \frac{d\rho^*_+}{d\langle k \rangle} = \frac{
    \Big(-\beta_1(1-\eta) - \beta_2 \eta \Big) 
        \Big[
            \sqrt{ {(\Lambda- \lambda_1- \lambda_2)}^2
		      -4(\Lambda + \tilde{\Lambda})(1-\lambda_1)}
            + (\Lambda- \lambda_1- \lambda_2)
        \Big]
        + 2(\Lambda+\Tilde{\Lambda})\beta_1(1-\eta)
    }{
    2\mu (\Lambda + \Tilde{\Lambda}) \sqrt{ {(\Lambda- \lambda_1- \lambda_2)}^2
		-4(\Lambda + \tilde{\Lambda})(1-\lambda_1)}
    }.
\end{equation}
From the latter expression, one deduces:
\begin{equation}
    \begin{split}
        \frac{d\rho^*_+}{d\langle k \rangle} = 0 \Longleftrightarrow 
    &\lambda_1 \Big[
                \beta_2^2 \eta^2 + \beta_1 \beta_2 \eta (1-\eta)
                \Big]
    -\lambda_2 \Big[
                \beta_1^2 {(1-\eta)}^2 + \beta_1\beta_2 \eta (1-\eta)
                \Big]
    -{\Big[\beta_1(1-\eta) + \beta_2 \eta\Big]}^2 \\
    &- \Tilde{\Lambda} \beta_1^2 {(1-\eta)}^2 + \Lambda \beta_1 \beta_2 \eta (1-\eta) = 0
    \end{split}
    \label{eq:derrho}
\end{equation}
Recalling that $\lambda_1 = \beta_1(1-\eta) \frac{\langle k \rangle}{\mu}$ and $\lambda_2 = \beta_2 \eta \frac{\langle k \rangle}{\mu}$, we see that the first two terms of the r.h.s of Eq.~\eqref{eq:derrho} cancel each other and we are left with:
\begin{equation}
    \frac{d\rho^*_+}{d\langle k \rangle} = 0 \Longleftrightarrow g(\eta):= -{\Big[
                            \beta_1(1-\eta) + \beta_2 \eta
                            \Big]}^2
                           -\Tilde{\Lambda}(\eta)\beta_1^2 {(1-\eta)}^2
                           +\Lambda(\eta)\beta_1 \beta_2 \eta (1-\eta) = 0,
\end{equation}
where we have stressed the dependence of $\Lambda$ and $\Tilde{\Lambda}$ on $\eta$.
Notice that the latter expression is independent of $\langle k \rangle$. As an illustration, let us consider the parameters used in Fig. \ref{fig:rhostar_2} in the main document. For these parameters, the function $g$ vanishes at the particular value $\eta = \eta_{\text{cross}} \approx 0.4$ as can be seen in the right panel of Fig.~\ref{fig:Non_monotonicity}. We can thus conclude that for $\eta < \eta_{\text{cross}}$, the infection decreases as $\langle k \rangle$ increases (see left panel the curve corresponding to $\eta = 0.3$) while for $\eta > \eta_{\text{cross}}$, the opposite behavior holds true (see left panel the curve corresponding to $\eta = 0.5$), i.e., $\rho^*$ is an increasing function of the average connectivity. 

\clearpage

\section{Conditions for the presence of the crossing phenomenon}\label{app:lambda_cross}
\label{sec:crossingcond}
The aim of this section is to investigate in more details the crossing behavior observed in Fig. \ref{fig:influence_pi} in the main document. Let us start by recalling that stationary densities are solutions of
\begin{equation}
\label{eq:rho0}
\rho \left(a +b \rho+c \rho^2\right)=0\, ,
\end{equation}
where
\begin{eqnarray}
	a &=& -\mu +\beta_1 (1-\eta) \langle k\rangle\\
	b &=& - \beta_1 (1-\eta) \langle k\rangle -\beta_2 \eta \langle k\rangle+\gamma_1\tau_3+\frac{\gamma_2\tau_2}{3}+\frac{2\gamma_3\tau_2}{3}\\
	c &=& -\gamma_1\tau_3 -\frac{\gamma_2\tau_2}{3}-\frac{2\gamma_3\tau_2}{3}-\frac{\gamma_4}{3}\tau_1-\gamma_5\tau_0\, .
\end{eqnarray}
Let us observe that $c<0$ for all values of the parameters, while $a$ and $b$ can change signs.

Let us recall the definitions $\Lambda := \Lambda_1+\Lambda_2 + \Lambda_3$ and  $\tilde{\Lambda}:=\Lambda_4 + \Lambda_5$ with:
\begin{equation}
\Lambda_1 = \frac{\gamma_1\tau_3}{\mu} \,
, \Lambda_2 = \frac{\gamma_2\tau_2}{3\mu} \, 
, \Lambda_3 = \frac{2\gamma_3\tau_2}{3\mu}, \\
\Lambda_4 = \frac{\gamma_4\tau_1}{3\mu}\, 
,\Lambda_5 = \frac{\gamma_5\tau_0}{\mu}\,\\
,\lambda_1 = \frac{\beta_1 \langle k \rangle}{\mu}\,
,\lambda_2 = \frac{\beta_2 \eta \langle k \rangle}{\mu}\, .
\end{equation}
and assume the number of triangles with a given number of positive/negative signs are given by
\begin{equation}
\tau_{0}=p_0\eta^3\langle k_\Delta\rangle\, , \tau_1=3p_1\eta^2(1-\eta)\langle k_\Delta\rangle\, ,\tau_{2}=3p_2\eta{(1-\eta)}^2\langle k_\Delta\rangle\, ,\tau_3=p_3{(1-\eta)}^3\langle k_\Delta\rangle \, .
\end{equation}
By dividing~\eqref{eq:rho0} by $\mu$, we get $a/\mu = -1+\lambda_1 (1-\eta)$ and thus rewrite the non zero solutions $\rho^*_\pm$ as follows
\begin{equation}
\lambda_1 (1-\eta) (1-\rho^*_\pm) = (\rho^*_\pm)^2(\Lambda + \tilde{\Lambda}) + \rho^*_\pm(\lambda_2 - \Lambda) + 1\, ,
\end{equation}
Let us notice that $\rho^*_\pm=1$ would imply $\tilde{\Lambda} + \lambda_2 + 1=0$ which can never happen being $\tilde{\Lambda}$ and $\lambda_2$ positive numbers. Hence, we can write (assuming $\eta \neq 1$):
\begin{equation}
\lambda_1\left(\rho^*_\pm\right) = \frac{
	(\rho^*_\pm)^2(\Lambda + \tilde{\Lambda}) + \rho^*_\pm(\lambda_2 - \Lambda) + 1
}{
	(1-\rho^*_\pm) (1-\eta)
}
\label{eq:lambda1}
\end{equation}
Let us observe that $\lambda_1$ has a vertical asymptote for $\rho^*_\pm =1$; being this function invertible close to $\rho^*_\pm =1$, we can conclude that $\rho^*_\pm \rightarrow 1$ for $\lambda_1\rightarrow \infty$.

To have the crossing property, we need to find $0<p_0^{(1)} \neq p_0^{(2)}$ such that $\lambda_1(\rho^*_\pm)\vert_{p_0^{(1)}} = \lambda_1(\rho^*_\pm)\vert_{p_0^{(2)}}$ for some $0<\rho^*<1$ (see Fig.~\ref{fig:influence_pi}); a straightforward computation allows to show that this is equivalent to require:
\begin{equation}
\label{eq:condrho1}
\rho^*_\pm = \frac{
	\phi_1 - \phi_2
}{
	c_2 - c_1
}\, ,
\end{equation} 
here $\phi_{1}:= \lambda_2 - \Lambda\vert_{p_0^{(1)}}$  and $c_{1} := \Lambda+ \tilde{\Lambda} \vert_{p_0^{(1)}}$ and similarly for $\phi_{2}$ and $c_{2}$.
Let us now consider the case of $\eta=\frac{1}{2}$ as in Fig.~\ref{fig:influence_pi} and let us assume $p_2=1$ and $p_1=p_3$. Recalling that $\tau_0 + \tau_1 + \tau_2 + \tau_3 = \langle k_\Delta \rangle$, we deduce that:
\begin{equation}
    p_0 = 5-4p_3.
\end{equation}
Taking into account the latter condition, one can show upon simplification that:
\begin{equation}
    \rho^*_{\pm} = \frac{\gamma_1}{\gamma_1 + \gamma_4 - 4\gamma_5}.
\end{equation}
Notice that the latter expression is independent of $p_3$ and belongs to $(0,1)$ provided $\gamma_4 > 4 \gamma_5$, let us observer that this condition is always met when $\gamma_5=0$ and $\gamma_4 > 0$, as considered in our analysis. The corresponding value $\lambda_{\text{cross}}$ can be deduced from Eq.~\eqref{eq:lambda1}.

\end{document}